\documentstyle[12pt,psfig]{article} 
\def\beq{\begin{equation}} 
\def\eeq{\end{equation}} 
\def\bea{\begin{eqnarray}} 
\def\eea{\end{eqnarray}} 
\def\bq{\begin{quote}} 
\def\eq{\end{quote}}

\def\beqa{\begin{eqnarray}}  
\def\eeqa{\end{eqnarray}}  
\def\be{\begin{equation}}  
\def\ee{\end{equation}}  
\def\beq{\begin{equation}}    
\def\eeq{\end{equation}}

\def\pa{\partial}

\def\cs{{\cal S}} 
\def\co{{\cal O}} 
 
\def\bi{\begin{itemize}}  
\def\ei{\end{itemize}}

\def\mh{\hat{\mu}} 
\def\nh{\hat{\nu}}

\def\lc{{\cal L}} 
\def\var{\vartheta}

\def\gappeq{\mathrel{\rlap 
{\raise.5ex\hbox{$>$}} 
{\lower.5ex\hbox{$\sim$}}}} 
 
\def\lappeq{\mathrel{\rlap{\raise.5ex\hbox{$<$}} 
{\lower.5ex\hbox{$\sim$}}}} 
 
\begin{document} 
\pagestyle{empty} 
\begin{flushright} 
{CERN-TH/98-118\\ 
OUTP-98-37-P\\ 
hep-ph/9805377} 
\end{flushright} 
\vspace*{5mm} 
\begin{center} 
{\bf  
Five-Dimensional Aspects  
of $M$-Theory Dynamics and Supersymmetry Breaking 
} \\ 
\vspace*{1cm}  
John Ellis$^{a)}$, Zygmunt Lalak$^{a,b)}$, Stefan Pokorski$^{a,b)}$ 
and Witold Pokorski$^{c)}$ 
\\ 
\vspace{0.3cm} 
\vspace*{2cm}   
{\bf ABSTRACT} \\ 
\end{center} 
\vspace*{5mm} 
\noindent 
We discuss the reduction of the eleven-dimensional $M$-theory 
effective  
Lagrangian, considering first compactification from eleven  
to five dimensions on a Calabi-Yau manifold, followed by reduction  
to four dimensions on 
an $S_1/Z_2$ line segment at a larger distance scale. 
The Calabi-Yau geometry leads to a 
structure of the five-dimensional Lagrangian that has more 
freedom than the eleven-dimensional theory. 
In five dimensions one obtains a non-linear $\sigma$ model coupled to 
gravity, which implies non-trivial dynamics for the scalar moduli fields  
in the bulk of the $Z_2$ orbifold. We discuss solutions to the  
five-dimensional equations of motion in the presence of sources localized  
on the boundaries of the $Z_2$ orbifold 
that may trigger supersymmetry breaking, e.g., gaugino 
condensates. The transmission of 
supersymmetry breaking from the hidden wall to the visible wall 
is demonstrated in specific  
models. The r\^ole of the messenger of supersymmetry breaking may be 
played by the gravity supermultiplet and/or by scalar hypermultiplets. 
The latter include the universal 
hypermultiplet associated with the Calabi-Yau 
volume, and also the hypermultiplets associated with deformations of 
its complex structure, which mix in general.   
  
\vspace*{1cm}  
\noindent 
 
\rule[.1in]{16.5cm}{.002in} 
 
\noindent 
$^{a)}$ Theory Division, CERN, Geneva, Switzerland.\\ 
$^{b)}$ Institute of Theoretical Physics, Warsaw University.\\ 
$^{c)}$ Department of Theoretical Physics, University of Oxford.\\ 
\vspace*{0.5cm} 
\begin{flushleft}  
CERN-TH/98-118\\ 
OUTP-98-37-P\\ 
May 1998 
\end{flushleft} 
\vfill\eject 

\setcounter{page}{1} 
\pagestyle{plain} 
 
\section{Introduction} 
 
The most plausible framework for a Theory of Everything (TOE) 
is generally agreed to be the theory formerly known as strings, 
presumably formulated in a suitable non-perturbative manner, 
termed $M$ theory. In this framework, the string coupling 
becomes a dynamical field that may be interpreted as an 
extra spatial dimension. When considered in the strong-coupling 
limit of the traditional ten-dimensional $E_8 \times E_8$ 
heterotic string~\cite{wh1, strw, wh}, $M$ theory appears able  
to reconcile the 
bottom-up estimate of the grand unification scale based on 
low-energy data from LEP and elsewhere with the top-down 
calculation of the string unification scale based on the 
Planck mass of the effective  
four-dimensional gravity~\cite{strw, bd, sstieb, li, john}.  
In this strong-coupling 
limit, the additional eleventh dimension becomes large 
compared with the four-dimensional Planck length. More detailed 
estimates suggest that it may even be considerably larger than 
the length scale of grand unification and the distance scale 
on which six internal dimensions are compactified~\cite{strw,bd}.  
 
The prototype formulation of eleven-dimensional $M$ theory that has been 
studied most extensively in the literature has been that in which  
the large eleventh dimension has the topology of 
an $S_1 / Z_2$ line segment of length $\pi \rho$, with two 
ten-dimensional walls located at its ends.  
This yields a 
strong-coupling limit of the traditional $E_8 \times E_8$  
heterotic string in which one (hidden) $E_8$ factor lives on the 
ten-dimensional wall at one end of the segment, with the other 
(observable) $E_8$ factor living on the opposite wall. 
Supersymmetry then requires that the bulk and the boundary hyperplane  
theories are not independent. An effective low-energy theory in  
five dimensions may be obtained by compactification of the six internal 
dimensions on a small 
Calabi-Yau space of size $R_{CY} \ll \rho $, which is capable of reducing 
the observable-sector gauge group to some subgroup of $E_6$, with 
$E_8$ or a proper subgroup on the hidden wall.  
The important aspect of this procedure  
is that the generalized Bianchi  
identity is satisfied only in the global sense, through the interplay  
of the sources on both walls.  
The presence of sources localized on the fixed planes means that in  
the subsequent dimensional reduction to four dimensions one has to take  
consistently into account the variation of the  
bulk fields across the 11th dimension 
\cite{strw,bd,dudas,nom,ovrut1,laltom,low5}. 
 
The systematic reduction of the eleven-dimensional $M$-theory Lagrangian  
to four dimensions has not yet been fully exploited. Since, as described 
above, 
$R_5 \gg R_{CY}$, this reduction should proceed in two steps.  
First the reduction from eleven to five dimensions should be performed.  
The structure  
of the Lagrangian in five dimensions is richer than in eleven dimensions,  
due to effects related to the geometry of the compact 
Calabi-Yau manifold. In five dimensions, one obtains a non-linear $\sigma$ 
model coupled to  
gravity~\cite{sier, sgun}, which implies non-trivial dynamics of scalar fields in the bulk 
of the $Z_2$ orbifold. The Calabi-Yau compactification also yields 
conventional four-dimensional gauge sectors on both walls, with specific 
couplings to the five-dimensional bulk theory. 
 
The complete derivation of the effective five-dimensional supergravity 
theory~\cite{ccf,aft,sharpe,low5} 
is not a straightforward task, as it has to take into account the  
above-mentioned solution of the eleven-dimensional Bianchi identities 
and the couplings to the gauge fields on the walls.  
As  
indicated by earlier investigations in strings and recently pointed  
out in the interesting paper~\cite{low5}, gauged extended  
supergravities are relevant for the low-energy description of 
$M$ theory. In contrast to ungauged supergravity theory, 
gauged supergravities in either five or four dimensions contain a 
generalized $D$-term potential for the scalar fields,  
which may contribute by itself 
to supersymmetry breaking~\cite{sgun, 
andr}. However, 
in the present paper we use the simple ungauged  
$N=2$ five-dimensional supergravity theory,  
with couplings derived from $M$ theory via Calabi-Yau 
compactification~\cite{ccf,aft,sharpe}, 
as the (toy) model for our discussion.  
The Bianchi identities are replaced in this model by additional 
sources, whose  
origin may be condensation of the gauge field strength and curvature 
tensors  
along the compact dimensions. These provide wall sources 
in the equation of motion for the universal  
$Z_2$-even field $S$, whose real part represents the volume of the 
Calabi-Yau space, which varies along the fifth dimension. 
 
The questions we want to ask in this paper are well defined 
already in this simplified setup, and we believe that the key 
answers will also be valid in the framework of gauged 
supergravity~\cite{low5}. 
We focus our attention on  
one important possible origin of sources localized on a boundary of the  
$Z_2$ orbifold, namely supersymmetry breaking in the hidden sector.  
If this occurs 
dynamically via gaugino condensation~\cite{revs},~\cite{msus}, we expect a  
coupling of the wall condensate to the bulk fields. 
These sources would then lead to non-zero modes in the solutions of the 
five-dimensional  
equations of motion, which have to be taken carefully into account  
in the construction of the effective four-dimensional Lagrangian.    
 
At this point arises the  
important question of the scale of the formation of the condensate, 
which we shall discuss  in more detail later. 
The original formulation by Horava~\cite{horava} is based 
on the assumption that the condensate  
forms already 
in eleven dimensions, perhaps due to a hidden-sector gauge group 
that is strongly coupled already very close to the eleven-dimensional  
Planck scale $m_{11}$. 
Noting a   
difference in the way a boundary gaugino condensate  
would enter the  
supersymmetric variations of fermions in the eleven-dimensional $M$-theory 
Lagrangian and in  
the ten-dimensional Einstein-Yang-Mills Lagrangian, he concludes that in 
the eleven-dimensional picture the condensate does allow  
{\em locally} for the existence of a spinor, giving vanishing  
supersymmetric variations of all fermions. However,   
this spinor is illegal from the point of view of the full model living on  
the $Z_2$ orbifold, since it is 
discontinous on the visible wall.  
This is a global obstruction precluding  
unbroken supersymmetry in the presence of such a `hard'  
eleven-dimensional condensate. 
 
The phenomenological hurdle to be crossed if supersymmetry breaking is due 
to a condensate forming close to the Planck scale  
is that of understanding the dynamical generation of any lower 
mass scale hierarchically smaller than the Planck scale. From 
this point of view, a 
more appealing situation would be if the condensate forms, as 
often postulated in the earlier days of string phenomenology, 
at a lower scale $\sim 10^{13}$ GeV. However, this would be 
below the apparent grand unification scale, and 
also below the scale of Calabi-Yau compactification of the internal six  
dimensions. Hence, in this more palatable case the condensation would actually  
occur in the context of en effective five- or even four-dimensional 
theory. 
 
One should stress that the two-step dimensional reduction, $11 \rightarrow 5 
\rightarrow 4$ dimensions, taking proper account of the  
dynamics in five dimensions, is a necessary 
framework  
for treating any of these three scenarios of the  
condensate formation. However, they would differ in the exact form of the 
sources  
on the hidden wall. In the first case, the source is determined by the  
reduction $11 \rightarrow 5$ of a stiff condensate that already exists in  
eleven dimensions. In the second one, it is legitimate to perform first  
the  
reduction of the supersymmetric model from eleven to five dimensions and 
then to supplement  
it with a stiff condensate source in five dimensions. Finally, in  
the third case a stiff condensate should be replaced by an effective  
boundary superpotential in five dimensions.     
 
The vacuum selection of the five-dimensional $\sigma$ model and the 
transmission of supersymmetry  
breaking to the observable wall are determined by the coupling 
of the $\sigma$ model describing the interactions of the bulk moduli  
to the boundaries, in the presence of sources on the hidden wall.  
The issue 
has already been studied in a toy model with rigid 
supersymmetry, consisting of a 
vector hypermultiplet in the five-dimensional bulk coupled to 
conventional four-dimensional chiral gauge theories on 
the walls~\cite{peskin, sharpe}. However, 
it is known that making  
five-dimensional supersymmetry local imposes a rather specific 
pattern of hypermultiplet couplings. It therefore seems 
opportune to revisit the mechanism for transmitting 
supersymmetry breaking, incorporating the general features of 
five-dimensional supergravity as well as the specific constraints 
that are imposed by Calabi-Yau compactification. 
 
As we recall in more detail in Section 2 below,  
five-dimensional supergravity theory contains in general a 
graviton supermultiplet, within which there is a single graviphoton 
vector state, a number of vector supermultiplets whose couplings 
are determined by a holomorphic trinomial, and a number of 
scalar hypermultiplets which parametrize a quaternionic 
manifold.  
A characteristic feature of the ungauged five-dimensional supergravity  
is the complete factorization 
of the manifolds parametrized by the scalars in the vector and 
scalar supermultiplets. This means, in particular, that the 
scalar hypermultiplet fields do not have these vector interactions\footnote{ 
Though the universal hypermultiplet does interact with the graviphoton in 
the gauged supergravity theory~\cite{low5}.}. 
 
In the case of compactification on a Calabi-Yau manifold,  
as also set out in Section 2, 
its topological numbers determine the numbers of 
vector and scalar supermultiplets: $n_V = h_{1,1} - 1, \; 
n_S = h_{2,1} +1$. Moreover, the trinomial characterizing 
the vector couplings is related to the intersection form 
of the Calabi-Yau manifold, there is a universal scalar 
hypermultiplet related to the volume of the Calabi-Yau 
manifold, and the geometry of the remaining scalar 
hypermultiplets is related to complex deformations 
of the Calabi-Yau manifold. 
The invariance of the eleven-dimensional supergravity theory 
compactified on an $S_1 / Z_2$ line segment under the $Z_2$  
symmetry of the orbifold, interpreted as a constraint on the fields  
present in the model formulated in the `upstairs' picture which we 
use also in five dimensions, determines through the  
Calabi-Yau compactification certain parity 
properties on the five-dimensional fields which we discuss at the end of  
Section 2. 
 
All Calabi-Yau compactifications yield a universal  
scalar hypermultiplet, but the properties of the other 
hypermultiplets associated with the complex structure 
moduli are quite model-dependent. We develop in Section 3 
a simple model with a single non-universal hypermultiplet, that 
serves to illustrate our subsequent discussion. 
 
In Section 4 we discuss the various possible scenarios for 
supersymmetry breaking mentioned above, that differ in the scale at which 
it is supposed to originate. We explore in most detail the case 
where this occurs between the Calabi-Yau and $S_1 / Z_2$ compactification 
scales, which we denote by $R_{CY}$ and $R_5$ respectively, paying close 
attention to the possible mechanisms 
for transmission of the supersymmetry breaking across the five-dimensional 
bulk. 
The $Z_2$ parity and Lorentz properties  
of the different fields tell us which must vanish and 
which may 
have non-zero expectation values on the walls. As discussed 
by~\cite{peskin}, the latter are essential for the possible 
transmission of supersymmetry breaking across the 
five-dimensional bulk. As we discuss in Section 4, the 
combination of these parity properties and four-dimensional 
Lorentz invariance implies that the vector supermultiplets 
may not transmit supersymmetry breaking directly, at least in the 
absence of Calabi-Yau deformation. However, this is 
possible via the gravity and matter hypermultiplets. In particular, the  
transmission via the hypermultiplets is 
modulated by the volume of the Calabi-Yau space. Another 
interesting feature is that, in general, 
the universal hypermultiplet mixes with the other hypermultiplets 
associated with complex deformations of the Calabi-Yau manifold, 
providing these with non-trivial dynamics. 
 
The five-dimensional equations of motion are used in Section 5 to  
find classical vacuum configurations for sources representing either 
a stiff condensate or a dynamical condensate, i.e., an effective boundary 
superpotential  
for $Z_2$-even moduli, in the simple Calabi-Yau model of Section 3. 
We explore the effects of the couplings between the different scalar 
fields in the non-linear $\sigma$ model in the five-dimensional bulk, and 
related non-linearities in the solutions of the equations of motion 
with boundary sources. The particular issues we study include the 
transmission of supersymmetry breaking from the hidden wall to 
the observable wall, and the choice of vacuum configuration. 
 
Finally, some perspectives for future work are outlined in Section 6. 
 
\section{Five-Dimensional Supergravity from Calabi-Yau Compactification} 
 
We first recall the general structure of the Lagrangian 
for  
ungauged $N=2$ Maxwell-matter supergravity in five dimensions~\footnote{We 
use the conventional term  
$N=2$, since the symplectic-Majorana gravitini transform as a doublet of 
the automorphism group of the supersymmetry algebra in five 
dimensions.}, 
especially its hyperplet structure and the geometry  
of the non-linear $\sigma$ model parametrized by the bosonic  
fields~\cite{sier, sgun}. 
We treat the fermions as 
symplectic-Majorana fields, i.e., there is an even number of  
them, forming conjugated pairs: 
$\lambda^a = \Gamma^5 (\lambda_a)^*$, where $\lambda^a=\epsilon^{ab} 
\lambda_b$ and $\lambda_a=\epsilon_{ba} \lambda^b$, with  
the totally antisymmetric two-index tensor $\epsilon$ 
defined so that $\epsilon^{12}=\epsilon_{12}=1$.  
As is well known the five-dimensional Lagrangian contains a 
gravity supermultiplet that includes 
the graviton, two gravitini $\psi^{\mu}_i$ and a graviphoton, 
vector supermultiplets that include gauge fields $A_{\mu}^A$, 
spin-1/2 fermions $\lambda^a_i$ and spin-0 fields $t^A$, 
and scalar hypermultiplets that include fermions $\lambda_b$ and 
spin-0 fields $\sigma^x$. It is frequently convenient to combine the  
graviphoton and the other vector fields using the indices $I, J, ...$.  
The index $i=1,2$ labels supersymmetries and is raised 
and lowered by the two-dimensional epsilon symbol. 
The Lagrangian may be written in terms of these physical fields as 
\beqa 
S  
& = & \int  \sqrt{g} \; (- \frac{R}{2} -\frac{1}{2} \bar{\psi}^i_\mu 
\Gamma^{\mu \nu \lambda} {\cal D}_\nu \psi_{\lambda i} \nonumber \\ 
& + & \frac{1}{2} \pa t^A \pa t^B G_{AB} - \frac{1}{4}  
G_{IJ} F^I F^J + \sqrt{g^{-1}} \frac{1}{48} d_{IJK}  
\epsilon A^I F^J F^K - \frac{1}{2} 
\bar{\lambda}^{ia} \Gamma^\mu {\cal D}_\mu {\lambda}_{i}^a \nonumber \\ 
& - & \frac{1}{2}g_{xy} \pa \sigma^x \pa \sigma^y  - 
\frac{1}{4} 
\bar{\lambda}^{b} \Gamma^\mu {\cal D}_\mu {\lambda}_b \nonumber \\ 
& - & \frac{3 i}{8 \sqrt{6}} h_I \bar{\psi}^i_\mu 
( \Gamma^{\mu \nu \alpha \beta}  +  2 \delta^{\mu \alpha}  
\delta^{\nu \beta} ) \psi_{\nu i } F^I_{\alpha \beta} \nonumber \\ 
& - &  
\frac{i}{2}  \bar{\lambda}^{ia} \Gamma^\mu \Gamma^\nu \psi_{\mu i } 
f^a_A \pa_\nu t^A  +  
\frac{1}{4} h^a_I \bar{\lambda}^{ia} \Gamma^\mu \Gamma^{\alpha \beta} 
\psi_{\mu i } F^I_{\alpha \beta}  \nonumber \\ 
 & + &  
\frac{i}{2}  \bar{\lambda}^{b} \Gamma^\mu \Gamma^\nu \psi_{\mu i } 
f^{ib}_x \pa_\nu \sigma^x  
 +  \frac{i}{2 \sqrt{6}} (\frac{1}{4}  
\delta_{ab} h_I + T_{abc} h^c_I )  
\bar{\lambda}^{ia}  
\Gamma^{\alpha \beta} {\lambda}_{i}^b F^I_{\alpha \beta} \nonumber \\ 
 & + & \frac{i \sqrt{6} }{32} h_I \bar{\lambda}_b  \Gamma^{\alpha \beta} 
 {\lambda}^b F^I_{\alpha \beta} \; + \; {\rm({four-fermion} \; terms)} 
\label{full5} 
\eeqa   
where the first three lines correspond to the gravity, 
vector and scalar supermultiplets, respectively. By $F^I$ we denote  
abelian field strengths of abelian vector bosons which can be  
present in the theory (at least one - the graviphoton - is always present).    
We recall that the 
couplings of the vector supermultiplets are characterized by a trilinear 
function ${\cal V}(X)$ where by $X$ we denote the scalar components of  
vector multiplets including the  
graviphoton, with Chern-Simons terms defined by 
\beq 
{\cal V}(X) = {1 \over 6} d_{IJK} X^I X^J X^K 
\eeq 
The coefficients forming a totally symmetric object $d_{IJK}$ must be  
constant, i.e. independent of the fields  
$t^A$, and all the other couplings in the vector Lagrangian are  
expressible in terms of them.  
The real scalars $t^A$ live on the 
hypersurface ${\cal V}(t) = 1$, and the kinetic-term metrics 
for the spin-1 and spin-0 fields are related as follows: 
\beqa 
G_{IJ} = - {1 \over 2} \pa_I \pa_J {\rm ln} {\cal V}|_{{\cal V} = 1}  
\nonumber \\ 
G_{AB}= G_{IJ} \pa_{t^A} X^I \pa_{t^B} X^J |_{{\cal V} =1}   
\label{metrics} 
\eeqa 
The metric $g_{xy}$ of the scalar hypermultiplets $\sigma^x$ is that 
of a quaternionic manifold, and the symbols $f^{ib}_x$ are the 
vielbeins 
of the metric $g_{xy}$ on the quaternionic manifold. 
It is noteworthy that the geometries of the spin-0 fields in the 
vector and scalar hypermultiplets are completely independent.  
This has the important phenomenological consequence that the scalar 
hypermultiplets can have no gauge interactions in ungauged $N=2$  
supergravity\footnote{Some are possible in the gauged versions,  
like those with 
the graviphoton in the gauged supergravity~of~\cite{low5}.}. 
 
We shall need for our subsequent analysis the supersymmetry 
transformation laws for the various hypermultiplets. These are 
\beqa 
\delta e^m_\mu & = & \frac{1}{2} \bar{\epsilon}^i \Gamma^m  \psi_{\mu i} 
\nonumber \\ 
\delta \psi_{\mu}^{ i} & = & ({\cal D}_\mu \epsilon)^i - \omega_{xj}^i 
( \delta \sigma^x ) \psi^j_\mu + \frac{i}{4 \sqrt{6}} h_I (\Gamma_\mu^{ 
\alpha \beta} - 4 \delta^\alpha_\mu \Gamma^\beta) \epsilon^i F^I_{\alpha \beta} 
\label{gtr} 
\eeqa 
in the case of  
the gravity supermultiplet, 
\beqa 
\delta A^I_\mu & = & - \frac{1}{2} h^I_a \bar{\epsilon}^i \Gamma_\mu  
\lambda^a_i + \frac{i}{4} \sqrt{6} h^I \bar{\psi}^i_\mu \epsilon_i  
\nonumber \\ 
\delta \lambda^a_i & = & - i f^a_A (\not\!\pa t^A) \epsilon_i -  
\omega_A^{ab} (\delta t^A) \lambda^b_i + \frac{1}{4} h^a_I  
\Gamma^{\mu \nu} F^I_{\mu \nu} \epsilon_i \nonumber \\ 
\delta t^A & = & \frac{i}{2} f^A_a \bar{\epsilon}^i \lambda^a_i  
\label{trvec} 
\eeqa 
for the vector supermultiplets and the graviphoton, and  
\beqa 
\delta \lambda^b & = & - i f^{ib}_x ( \not\!\pa \sigma^x) \epsilon_i 
 - \omega_x^{b\;c} ( \delta \sigma^x) \lambda^c \nonumber \\ 
\delta \sigma^x & = & \frac{i}{2} f^x_{i b} \bar{\epsilon}^i \lambda^b 
\label{trhyp} 
\eeqa 
for the scalar hypermultiplets. 
 
After summarizing the basic facts about general ungauged  
$N=2$ d=5 matter-Maxwell  
supergravity, we would like to restrict  ourselves to specific models  
generated through compactifications on Calabi-Yau spaces. More precisely,   
following~\cite{strw, wh1, wh}, we take as a starting point 
eleven-dimensional supergravity coupled to ten-dimensional  
supersymmetric Yang-Mills  theories living on two separated 
boundary walls. If one  makes a field-theoretical compactification  
on a Calabi-Yau threefold, one obtains a bulk 
five-dimensional supergravity theory.  
We expect that this construction should lead to  
a model that can be described 
within the general framework (\ref{full5}), but  
with specific constraints due to its origin in  
a Calabi-Yau compactification of eleven-dimensional supergravity. 
These should include constraints on the number of 
vector hypermultiplets $A^A_{\mu}$ and the trilinear geometrical 
function ${\cal V}(t)$ that characterizes their self-couplings, as 
well as constraints on the number of scalar hypermultiplets and 
constraints on the quaternionic manifold that describes their 
geometry~\cite{ccf,aft}.  
 
Before entering into more detail on these subjects, we 
recall that the Calabi-Yau manifold is expected to be deformed \cite{strw}.  
This is because  
the generalized Bianchi identity in eleven dimensions is fullfilled  
only globally, through the interplay between non-zero sources  
located on both walls, which leads to a 
non-zero antisymmetric-tensor field  background interpolating across the 
eleventh 
(to be renamed fifth) dimension between these sources.   
This non-zero background, combined with the requirement of the  
vanishing supersymmetry variation of one   
gravitino on the walls, 
leads to a non-vanishing correction to the metric. The simplest 
consequence of this is a linear variation of 
the Calabi-Yau volume along the $S^1 /Z_2 $ line segment.  
 
Among the metric deformations, one may choose to restrict oneself  
to such deformations which are independent of the Calabi-Yau coordinates  
and depend only on $x^5$. This leaves the compact six  
dimensions a Calabi-Yau space for each $x^5$. 
Then, the natural first step towards the realistic compactification of  
$M$ theory down to five dimensions is to proceed 
with the standard  
Calabi-Yau compactification~\cite{aft, ccf}.  
The assumption made above has the useful consequence that the corrections to 
the  
metric are taken into account as $x^5$-dependent configurations of the  
bulk moduli fields. However, it should be clear that, as the bulk 
moduli  
are part of the non-linear $\sigma$ model in five dimensions, their vacuum  
contains more structure than a simple reconstruction of the original  
deformation computed in eleven dimensions. This is consistent with  
the observation made in the forthcoming sections of this paper that the actual  
vacuum configurations of fields do not have simple linear dependences on  
$x^5$. 
 
In five dimensions, where we have at our disposal only fields which are  
true zero modes from the point of view of the Calabi-Yau space, the 
obvious question is  
how the non-zero background of the even modulus $S$, representing the 
Calabi-Yau volume,  
can get excited. As we have said before, in eleven dimensions the source of  
the deformation of the volume is the non-zero antisymmetric-tensor field  
background. In five dimensions, the way to produce the non-zero slope in a  
field is to couple it to  sources located on the boundaries. These sources  
induce the expected dependence of the volume on the fifth coordinate 
$S(x^5)$, and so represent the non-zero antisymmetric-tensor  
background in the effective five-dimensional Lagrangian.  
As explained in the  
forthcoming sections of this paper, we take such a background 
simply into account by  
assuming additional $\delta$-function sources, located on the walls, in 
the equation of motion for the field $S$, whose consequences will  
be discussed later. These additional sources can in principle be 
incorporated 
in the effective five-dimensional Lagrangian. For the purpose of this 
paper we leave them at the level of equations of motion\footnote{However,  
we mention that  
these corrections  
underlie the gauged five-dimensional supergravity 
in~\cite{low5}}.  
 
Calabi-Yau compactification of $M$ theory yields a 
five-dimensional supergravity lagrangian of the general 
form ({\ref{full5}), with specific geometrical constraints 
related to the topological structure of the Calabi-Yau 
manifold. The indices $A,\, a$ are  
$O(h(1,1)-1)$  
vector indices, and $I,J$ run over the  
range $1 \ldots h_{1,1}$. The index $b$ on the fermions belonging to the 
scalar hypermultiplets  
runs over the range $1 \ldots 2(h_{2,1}+1)$ and transforms as  
the fundamental representation of $USp(h_{2,1}+1)$. The index $x$, 
that counts the real bosonic degrees of freedom in hypermultiplets, 
runs over the range 
$1 \ldots 4 (h_{2,1}+1)$. These matter fields, when coupled to  
supergravity, form a quaternionic manifold of real dimension  
$4 (h_{2,1}+1)$, as discussed in the general case of matter coupled to 
$N=2$ supergravity~\cite{bagger, sier, andr}.  
The tangent-space metric of this quaternionic manifold is the invariant 
antisymmetric  
matrix of the group $USp(1) \times USp(h_{2,1}+1)$. The vielbein 
$f^a_A$ of the metric of the manifold spanned by the scalar 
components of the 
vector superplets may be found using the $SO(h_{1,1}-1)$ tangent group.  
The corresponding connections may be formed out of the vielbeins in the  
standard way, and we denote them by $\omega^i_{j}$, 
$\omega^a_{b}$,$\ldots$, respectively.  
 
The reduction of the supersymmetry on the boundary walls to N=1 is  
accomplished, exactly as in the original  
eleven-dimensional theory, by representing the $Z_2$ geometrical symmetry 
of the $S^1 / Z_2$ orbifold on the fields. Then the $Z_2$-odd components 
are 
projected out at the boundaries, leaving a number of  
degrees of freedom corresponding to simple chiral $N=1$ supersymmetry on  
these four-dimensional hyperplanes.   
In the upstairs picture which we adopt in this paper,  
this means that we define the action of the $Z_2$ on the bosonic and 
fermionic  
fields in the model so as to leave the five-dimensional action  
invariant\footnote{In principle, this $Z_2$ symmetry could be embedded 
non-trivially in the symmetry group of the underlying  
Calabi-Yau space, which would lead to  
interesting consequences. However, for the purpose of the  
present paper we restrict ourselves to the case where 
the $Z_2$ acts trivially on Calabi-Yau geometry.}. We discuss 
the $Z_2$ transformation properties of the bulk five-dimensional 
fields in more detail at the end of this section. 
 
We define the Calabi-Yau metric moduli through~\cite{bodner}  
\beqa 
i \delta g_{i \bar{j}}& = & \sum^{h_{1,1}}_{A=1}  
\delta M^A V^A_{i \bar{j}}, \; \delta g_{ij} = \sum^{h_{2,1}}_{\alpha=1}  
\delta \bar{Z}^{\alpha} \bar{b}_{\alpha i j}  
\eeqa 
where the $M^A$ are real moduli corresponding to the deformations of the  
K\"ahler class, and the ${Z}^{\alpha}$ are complex moduli corresponding to  
the deformations of the complex structure of the Calabi-Yau manifold,  
the $V^A_{i \bar{j}}$ are the harmonic $(1,1)$ forms, and  
the $\bar{b}_{\alpha i j}$ are harmonic $(2,1)$ forms.  
In addition, we denote the holomorphic  
$(3,0)$ form by $\Omega_{ijk}$.  
 
The kinetic-energy terms for the 
metric moduli can be obtained from the reduction of the 
eleven-dimensional Einstein-Hilbert  
action~\cite{ccf,aft}. The resulting parts of the five-dimensional action 
are 
\beq 
S_g = \int \sqrt{g^{(5)}} ( \frac{1}{2} \pa M^A \pa M^B  V  
\hat{G}_{AB} - V \frac{R^{(5)}}{2} + V G_{\alpha \bar{\beta}} 
\pa Z_{\alpha} \pa \bar{Z}^{\bar{\beta}}) 
\eeq 
where $V$ is the volume of the Calabi-Yau three-fold and $g^{(5)}, 
R^{(5)}$, etc., denote gravitational quantities in five dimensions.  
In order to identify correctly the degrees of freedom and the 
hypermultiplets to which they belong, one must subtract the volume modulus  
from the real moduli of the $(1,1)$ type. To this end, we define 
real fields $t^A$ such that $M^A = t^A V^{1/3}$ and ${\cal V}(t^A) =1$. 
In terms of $t^A, V, Z^{\alpha}$, and after a suitable Weyl  
rescaling of the metric, we obtain 
\beq 
S_g = \int \sqrt{g} ( \frac{1}{2} \pa t^A \pa t^B   
 G_{AB} -  \frac{R}{2} +  G_{\alpha \bar{\beta}} 
\pa Z_{\alpha} \pa \bar{Z}^{\bar{\beta}} - \frac{1}{2} (\pa \log V)^2 
 ), 
\eeq    
which has a canonical Einstein-Hilbert term. Denoting the 
Calabi-Yau manifold by $K$, one may express the metric $G_{\alpha 
\bar{\beta}}$ in the form 
\beq 
G_{\alpha \bar{\beta}} = - { i \over 4}  
{\int_K b_\alpha \wedge {\bar b}_{\bar \beta} \over V} 
\label{complexmetric} 
\eeq 
and we discuss shortly the form of $G_{AB}$. 
 
The eleven-dimensional antisymmetric tensor $C_{MNP}$ 
also yields massless fields in five dimensions, which 
are related to harmonic forms on the Calabi-Yau space.  
The components of $C_{MNP}$ which have all five-dimensional 
indices give rise, upon using Hodge duality and the tree-level 
equations of motion, to a single real scalar which we denote by $S_2$, 
the components with one five-dimensional index, one holomorphic and one 
antiholomorphic  
index give $h_{1,1}$ vector bosons in five dimensions, and the 
components  
with two holomorphic and one antiholomorphic index give $h_{2,1}$ 
complex scalars. Together with the fields coming from the  
reduction of the metric, including the five-dimensional graviton, these 
zero modes provide 
the bosonic components of the gravity hypermultiplet, 
$h_{1,1}-1$ vector multiplets, and $h_{2,1}+1$  
hypermultiplets. One of these scalar hypermultiplets contains in its 
bosonic sector 
the field $S=S_1 + i S_2$, where $S_1$ corresponds to the   
Calabi-Yau volume, and the complex scalar $C$ coming from the reduction of 
the  
antisymmetric tensor components with three antiholomorphic indices, 
namely $C_{ijk}$. This tensor is called the universal hypermultiplet, as 
it is always present in Calabi-Yau compactifications, even when the number  
$h_{2,1}$ of independent $(2,1)$ forms vanishes.  
 
The remaining hypermultiplets, containing complex scalars coming from the  
reduction of the metric and antisymmetric tensor field, are 
non-universal. Since the physics of hypermultiplets will be of importance 
later in this paper, we write down here for completeness the part of the 
five-dimensional Lagrangian  
which contains bosonic fields from the hypermultiplets, including 
the graviphoton, which is  
included with the other vector fields: 
\beq 
S_V  = \int ( \sqrt{g}    
 \frac{1}{2} \pa t^A \pa t^B G_{AB} +\sqrt{g} \frac{1}{4}  
G_{IJ} F^I F^J + \frac{1}{48} d_{IJK} \epsilon A^I F^J F^K) 
\eeq    
where $\epsilon$ is the completely antisymmetric tensor in 
five dimensions, and the real scalars  
$t^A$ live on the hypersurface ${\cal V}(t)=1$. 
One has the representation 
\beq 
d_{IJK}  =  \int_K V^I \wedge V^J \wedge V^K, \nonumber \\ 
\label{CYd} 
\eeq 
in terms of which the metrics $G_{AB}, G_{IJ}$ are then given by 
(\ref{metrics}). 
The expressions (\ref{CYd}) allow one to identify properly the 
constraints imposed by Calabi-Yau compactification 
of eleven-dimensional supergravity on the coefficients in 
the general five-dimensional supergravity  
Lagrangian (\ref{full5})~\cite{sier}.  
 
Before leaving this section, we discuss the $Z_2$ properties~\cite{dudas, 
sharpe,low5} 
of various fields in the five-dimensional Lagrangian. 
These  
can be read off the eleven-dimensional supergravity Lagrangian, in  
particular by demanding $Z_2$ invariance of the topological $C G G$ 
term. Denoting the Minkowski-space coordinates by $\mh,\nh$, etc.,  
the only odd components of the metric are  
$g_{\mu 5}, g_{i5}, g_{\bar{i}5}$, eg. $g_{\mu 5}(-x^5) =  
- g_{\mu 5} (x^5)$. The remaining metric components are $Z_2$ even. 
In the case of the antisymmetric tensor field $C_{MNP}$, the parity 
assignments are just the opposite:  
the components without the index $5$ are odd, and those with index $5$  
are even, e.g., $C_{\hat{\mu} i \bar{j}}(-x^5) = - C_{\mu i \bar{j}}(x^5)$ 
and $C_{5 i \bar{j}}(-x^5) =  C_{5 i \bar{j}}(x^5)$.  
Consequently, the moduli scalars $t,V,Z$ are all even. 
As for the vectors, their $A^I_{\mu}$ components are odd, but their 
components $A^I_{5}$ are even. The complex scalars coming from  
components of $C_{MNP}$ with all indices holomorphic or 
antiholomorphic are 
odd, and the remaining real scalar, coming from components  
with the index structure $C_{5 \mu \nu}$, is even.  
 
We define the $Z_2$ action on fermions as follows: 
\beqa 
\lambda^a (-x^5) &=& i \Gamma^5 \lambda_a(x^5) \nonumber \\ 
\psi^i_{\hat{\mu}} (-x^5)& = & i \Gamma^5 \psi_{i \hat{\mu}} (x^5)  
\nonumber \\ 
\psi^i_{5} (-x^5)& = & - i \Gamma^5 \psi_{i 5} (x^5) \nonumber \\ 
\epsilon^i (-x^5)&  = & i \Gamma^5 \epsilon_i (x^5) 
\label{zet} 
\eeqa 
The advantage of this definition of the action of $Z_2$  
is that it exchanges left (from the four-dimensional point of view) 
components 
of a pair of symplectic spinors between the partners, and similarly  
the right components. If the spinors $\lambda^a$, $a=1,2$, form a  
symplectic pair, then 
\beq 
\lambda^1_L \rightarrow i \lambda^2_L, \; \lambda^1_R  
\rightarrow -i \lambda^2_R 
\label{sympl} 
\eeq 
One should note that, in the representation of the charge conjugation 
matrix we have chosen, the left and right components of the  
symplectic spinors are related as follows 
\beq 
\lambda^1_L = (\lambda^2_R)^*, \; \lambda^1_R = - (\lambda^2_L)^* 
\label{zetl} 
\eeq 
Hence, for instance, one can choose to work with the left components  
of the symplectic spinors only, and these will contain the full  
information carried by a pair of symplectic  
spinors, and provide a nonsinglet 
representation for the $Z_2$.  
Now one can easily find the  
combinations of spinors which are eigenstates of the $Z_2$ parity: 
the combination $\lambda_L^+ = \lambda_L^1 + i \lambda_L^2$ is an even 
singlet, whilst 
$\lambda_L^- = i \lambda_L^1 +  \lambda_L^2$ is odd.  
Similarly, for the fermionic parameters of supersymmetry transformations, 
we see that 
$\epsilon_L^+ = \epsilon_L^1 + i \epsilon_L^2$ generates $Z_2$-even 
transformations, whilst $\epsilon_L^- = i \epsilon_L^1 + \epsilon_L^2$  
generates $Z_2$-odd ones. These properties will be important when we 
discuss 
supersymmetry on the three-branes. There, the $Z_2$-odd states and  
supersymmetries are projected out of the spectrum, whilst their  
derivatives with respect to the odd coordinate $x^5$ are $Z_2$ even, and  
may sneak in to play a r\^ole in the  
four-dimensional physics on the three-branes. 
 
\section{The Scalar Hypermultiplet Sector in a Prototype for Calabi-Yau 
Compactification of $M$ Theory} 
 
As we have already remarked, the 
general form of the moduli space in  
five-dimensional supergravity is a direct product $M=M_K \times Q$, 
where $M_K$ is a K\"ahler manifold and $Q$ is a quaternionic manifold. 
There is in general~\cite{subhar} a mapping 
between these two manifolds, called the $s$ map, which means 
that we only need to know 
the metric for the K\"ahler manifold characterizing 
the dynamics of the complex  
metric moduli of $(2,1)$ type to describe the geometry of  
the full moduli space. 
Unfortunately, despite the existence of this 
$s$ map, the general case of a phenomenologically-relevant Calabi-Yau 
space is very complicated to treat in detail.  
We therefore 
discuss a toy model~\cite{subhar}  
which is not phenomenologically appealing, as its 
Yukawa couplings vanish, 
but does allow us to write down a non-trivial, explicit 
Lagrangian for the hypermultiplets which may be used to 
discuss issues of phenomenological relevance, such as 
supersymmetry breaking.  
 
The moduli space for this Calabi-Yau space is 
given by the product of the K\"ahler manifold 
\beq 
M_K={SU(1,n) \over U(1) \times SU(n)} 
\eeq 
with the quaternionic manifold 
\beq 
Q={SU(2,n+1) \over SU(2) \times SU(n+1) \times U(1)}, 
\eeq 
where $n$ is the number of hypermultiplets in the theory: in our case, 
$n = h_{2,1}+1$, 
where $h_{2,1}$ is the Hodge number of the Calabi-Yau manifold. 
In the present case, the quaternionic manifold is in fact also 
K\"ahler, which enables us to write down an explicit 
K\"ahler potential for the moduli space: 
\beq 
K_m=K+\tilde{K}, 
\eeq 
where 
\beq 
K(Z, \bar{Z})= - \log \left( 2 (1- Z \bar{Z}) \right) 
\eeq 
and 
\bea 
\tilde{K}(Z,\bar{Z},S,\bar{S},C_0,\bar{C_0},C_1,\bar{C_1}) & = &  
  -\log ( S + \bar{S} - {(1 + Z \bar{Z}) (C_0 + \bar{C_0})^2  
\over 1 - Z \bar{Z}}+ \\ \nonumber 
 & + & 
{ 2 (Z + \bar{Z}) (C_1 + \bar{C_1}) (C_0 + \bar{C_0}) +  
        (1 + Z \bar{Z}) (C_1 + \bar{C_1})^2 \over 1 - Z \bar{Z}} ). 
\eea 
In what follows, we limit ourselves to the case where there is only  
one non-universal 
hypermultiplet, i.e., the Hodge number  
of the corresponding Calabi-Yau space is  
$h_{2,1} = 1$. 
In our notation, the pair of complex scalars  
$(S,C_0)$ belongs to the universal  
hypermultiplet, and the pair $(Z,C_1)$ to the single remaining  
non-universal hypermultiplet. 
 
Using the above K\"ahler potentials, we can construct  
the Lagrangian describing the dynamics of the scalar fields: 
\bea 
\label{lagra} 
e^{-1} {\cal L} &=& - K_{Z \bar{Z}} \partial_{\mu} Z \partial_{\mu} \bar{Z}  
 -  \tilde{K}_{S \bar{S}} \partial_{\mu} S \partial_{\mu} \bar{S}- 
\tilde{K}_{S \bar{Z}} \partial_{\mu} S \partial_{\mu} \bar{Z}  
 -   \tilde{K}_{Z \bar{S}} \partial_{\mu} Z \partial_{\mu} \bar{S}  
 -  
\tilde{K}_{S \bar{C}_i} \partial_{\mu} S \partial_{\mu} \bar{C}_i  \\ \nonumber 
& - & \tilde{K}_{C_i \bar{S}} \partial_{\mu} C_i \partial_{\mu} \bar{S}-  
\tilde{K}_{C_i \bar{C}_j} \partial_{\mu} C_i \partial_{\mu} \bar{C}_j  
 -  \tilde{K}_{C_i \bar{Z}} \partial_{\mu} C_i \partial_{\mu} \bar{Z}- 
\tilde{K}_{Z \bar{C}_i} \partial_{\mu} Z \partial_{\mu} \bar{C}_i 
\eea 
For comparison with the general formalism outlined above, and for our 
own further purposes, it is convenient to write the above Lagrangian 
in the form of the quaternionic metric $\sigma_{xy}$: 
\beq 
e^{-1} {\cal L} = ds^2=- g_{xy} \pa \sigma^x \pa \sigma^y. 
\eeq 
In this notation, the indices $x$ and $y$ run over all the fields in 
(\ref{lagra}), in the order $S, C_0, Z, C_1$. Since the 
exact metric $g_{xy}$ has a 
complicated expression, we limit ourselves to the limit where 
the fields $C_i$ and $Z$ are small, corresponding to a small 
deformation of the Calabi-Yau manifold.  
in this limit, we obtain the simplified expression 
\beq 
\label{metric} 
g = \left[ 
\begin{array}{cccc} 
{1 \over (S + \bar{S})^2} &  {{-2(C_0 +\bar{C_0})} \over  (S + \bar{S})^2}  
& 
0 &  {{-2(C_1 +\bar{C_1})} \over  (S + \bar{S})^2} \\ 
 {{-2(C_0 +\bar{C_0})} \over  (S + \bar{S})^2} & {2 \over S + \bar{S}} & 
 {{2(C_1 +\bar{C_1})} \over  S + \bar{S}} &  {{2(Z +\bar{Z})} \over  S  
+ \bar{S}} 
\\ 
0 &  {{2(C_1 +\bar{C_1})} \over  S + \bar{S}} & 1 &  {{2(C_0 +\bar{C_0})}  
\over  S + \bar{S}} \\ 
 {{-2(C_1 +\bar{C_1})} \over  (S + \bar{S})^2} &  {{2(Z +\bar{Z})} \over  
 S + \bar{S}} &  {{2(C_0 +\bar{C_0})} \over  S + \bar{S}} & 
{2 \over S + \bar{S}} 
\end{array} 
\right] 
\eeq 
when we expand the 
quaternionic metric up to linear order in all the fields except $S$, 
 
For our subsequent analysis of supersymmetry breaking, 
we need vielbeins corresponding to the metric 
$g_{xy}$, and also the part of the connection which corresponds to the  
$Sp(1)$ subgroup of the tangent group. We introduce the vielbeins  
in the following way 
\beq 
\label{viel} 
ds^2=V^T \bar{\Omega} V, 
\eeq 
where the $V$ are vielbeins with the components 
\beq 
V=\left( 
\begin{array}{c} 
u \\ 
v \\ 
w \\ 
r \\ 
-\bar{r} \\ 
\bar{w}\\ 
\bar{v}\\ 
-\bar{u} 
\end{array} 
\right), 
\eeq 
and with the metric 
\beq 
\bar{\Omega}=\left( 
\begin{array}{cccccccc} 
0 & 0 & 0 & 0 & 0 & 0 & 0 & -{1 \over 2} \\ 
0 & 0 & 0 & 0 & 0 & 0 & {1 \over 2} & 0 \\ 
0 & 0 & 0 & 0 & 0 & {1 \over 2} & 0 & 0 \\ 
0 & 0 & 0 & 0 & -{1 \over 2} & 0 & 0 & 0 \\ 
0 & 0 & 0 & -{1 \over 2} & 0 & 0 & 0 & 0 \\ 
0 & 0 & {1 \over 2} & 0 & 0 & 0 & 0 & 0 \\ 
0 & {1 \over 2} & 0 & 0 & 0 & 0 & 0 & 0 \\ 
-{1 \over 2} & 0 & 0 & 0 & 0 & 0 & 0 & 0  
\end{array} 
\right). 
\eeq 
Using the metric (\ref{metric}) and the definition (\ref{viel}), we 
find the following components of the vielbeins: 
\beq 
u=\sqrt{x} dS, 
\eeq 
\beq 
v={n \over \sqrt{x}} dS + \sqrt{y- {n^2 \over x}} dC_0, 
\eeq 
\beq 
w= \overline{{b \over  \sqrt{y- {n^2 \over x}}}} dC_0 + 
\sqrt{1 - {b^2 \over  | y- {n^2 \over x}|}} dZ, 
\eeq 
\beq 
r= {p \over \sqrt{x}} dS + \overline{ {q- {n p \over x} \over  
\sqrt{y- {n^2 \over x}}}} dC_0 + 
{a-{b (q^2 -{a p \over x}) \over | y- {n^2 \over x}|} 
\over \sqrt{1- {b^2 \over  | y- {n^2 \over x}|}}} dZ + 
\sqrt{y- {p^2 \over x}- {(q- {n p \over x})^2 \over  
|y- {n^2 \over x}|}- {\left( a-{b (q^2 -{a p \over x}) \over | y-  
{n^2 \over x}|} \right)^2 
\over |1- {b^2 \over  | y- {n^2 \over x}|}|}} dC_1, 
\eeq 
where 
\beq 
x={1 \over (S + \bar{S})^2}, 
\eeq 
\beq 
y={2 \over S + \bar{S}}, 
\eeq 
\beq 
n=-{2(C_0 +\bar{C_0}) \over (S + \bar{S})^2}, 
\eeq 
\beq 
p=-{2(C_1 +\bar{C_1}) \over  (S + \bar{S})^2}, 
\eeq 
\beq 
q={{2(Z +\bar{Z})} \over  S  
+ \bar{S}}, 
\eeq 
\beq 
a={{2(C_0 +\bar{C_0})} \over  S + \bar{S}}, 
\eeq 
\beq 
b={{2(S_1 +\bar{C_1})} \over  S + \bar{S}}. 
\eeq 
Suppressing $\sigma$-model indices $x$, 
the connections for this quaternionic manifold are defined as  
follows  
\beq 
d f^{ia} + \omega^i_{j} f^{ja} + \omega^a_{b} f^{ib} =0 
\eeq 
The parts of the $Sp(1)$ connection $\omega^i_{j}$ which are  
relevant for the projection  
on the boundary walls are  
\beqa 
\omega^1_{1} & = & \frac{1}{4} e^{\tilde{K}} (dS - d \bar{S}) +  
\frac{1}{4} \frac{\bar{Z} d Z + Z d \bar{Z} }{1-|Z|^2} + \ldots \nonumber \\ 
\omega^2_{2} & = & -\frac{1}{4} e^{\tilde{K}} (dS - d \bar{S}) - 
\frac{1}{4} \frac{\bar{Z} d Z + Z d \bar{Z} }{1-|Z|^2} + \ldots \nonumber \\ 
\omega^1_{2} & = & -2 e^{\frac{K + \tilde{K}}{2}} \bar{Z} d C + \ldots  
\nonumber \\ 
\omega^2_{1} & = & 2 e^{\frac{K + \tilde{K}}{2}} Z d \bar{C} + \ldots 
\eeqa  
These explicit formulae will be useful in the subsequent discussion 
of supersymmetry breaking.

\section{Scenarios for Supersymmetry Breaking in $M$ Theory} 
 
We now address the problem of supersymmetry breaking in $M$ theory 
using the general five-dimensional framework set out above. 
The appearance of two compactification scales, namely those 
of the Calabi-Yau manifold ($R_{CY}$) and of the $S_1/Z_2$ line segment 
($R_5$), with $R_{CY} < R_5$, 
enables us to distinguish three generic possibilities for the 
origin of supersymmetry breaking. It may originate either (a) in dynamics 
at a length scale less than $R_{CY}$, or (b) at a distance scale 
intermediate between $R_{CY}$ and $R_5$, or (c) at a distance scale 
larger than $R_5$. Examples of scenario (a) include the  
eleven-dimensional formulation of gaugino condensation and 
the Scherk-Schwarz mechanism~\cite{dudas,anton}. 
Scenario (b) may arise if  
there is supersymmetry breaking in the effective four-dimensional 
field theory on the hidden wall, e.g., as a result of strong gauge 
interactions that cause gaugino condensation~\cite{msus}.  
Scenario (c) would 
arise, e.g., if the hidden-sector gauge dynamics becomes strong only at a 
distance scale larger than $R_5$. In the following we discuss each of 
these scenarios from the point of view of the five-dimensional 
effective supergravity theory.  
 
We emphasize in advance 
that the intermediate five-dimensional supergravity stage is relevant in 
all these cases, because 
compactification on a Calabi-Yau space introduces new physics, 
to some extent compactification-dependent. This has an  
unavoidable r\^ole to play, as it couples directly to 
the trigger for supersymmetry breaking, which is a wall effect, 
and transmits dynamics between the walls. Hence, 
even in case  
(a), the steps to be taken include the compactification of the  
eleven-dimensional condensate on the Calabi-Yau manifold~\cite{laltom}, 
and the consideration of its  
physics in connection with that of the $\sigma$ model in the bulk.  
For the same reason, in case (c), before one integrates out  
the fifth dimension, one should  solve the five-dimensional equations of 
motion with the fermionic bilinear replaced by the effective  
superpotential for the even moduli~\cite{laltom}.  
The fifth  
dimension does not decouple in a trivial way for length scales  
larger than its radius.  
  
Let us first consider supersymmetry breaking at short distances. 
In this case, it is not evident that a field-theoretical 
description is adequate: one may need to use the whole 
paraphernalia of extended objects, including strings,  
membranes, etc.. Nevertheless, one may choose to 
investigate the consequences of postulating 
gaugino condensation within the field-theoretical 
supergravity framework. In such a scenario of  
supersymmetry breaking at short distances, the 
vacuum expectation value of a fermion bilinear would be `hard', 
in the sense of being independent of the moduli of 
compactification. If the field-theoretical description is adequate, 
then the analysis of Horava~\cite{horava} applies, and should be followed 
by compactification of the internal six dimensions, and subsequent 
analysis repeating the procedure described below for the case (b). 
 
Next we explore in detail the possibility (b) that supersymmetry breaking 
originates at a distance scale intermediate between $R_{CY}$ and 
$R_5$ (intermediate-scale supersymmetry breaking).  
In this case, we expect that the origin of 
supersymmetry breaking may be described by some effective 
four-dimensional field theory on the hidden wall. We do not 
discuss here the details how this may arise: strongly-coupled 
gauge dynamics leading to gaugino condensation may be one 
option, but 
there may well be others. In any such scenario, one must 
discuss how such supersymmetry breaking may be transmitted 
through the five-dimensional bulk to the observable wall. 
Here several issues arise, including the couplings of the bulk theory to 
the walls, whether the bulk theory admits supersymmetry breaking 
of the type advocated, and the relative magnitudes of the 
supersymmetry breaking at each end of the $S_1 / Z_2$ line segment. 
We now address each of these issues in turn. 
 
We have defined the bulk theory in the previous sections. However, this 
theory as it stands  
contains no potential for scalar fields, but consists of kinetic terms  
and derivative interactions only~\footnote{Although we note that  
potential terms are present in gauged N=2 supergravity~\cite{andr}, see also  
\cite{low5}.}.  
However, we know that compactification on a Calabi-Yau threefold 
generates non-derivative  
couplings on both walls, and in particular that non-perturbative  
effects may create a non-trivial  
potential for the even bulk moduli, which are legal chiral superfields  
on the walls. Let us assume that such sources are indeed present in  
the theory, and assume for the time being that they behave like a covariantly  
constant condensate, i.e., are proportional to the unique Calabi-Yau 
three-form  
$\Omega_{ijk}$, coupling directly to the fifth derivative of the  
$Z_2$-odd complex scalar in the universal hypermultiplet.  
A unique form for the five-dimensional coupling of the source to the 
universal  
hyperplet is suggested by the reduction of the eleven-dimensional  
`perfect-square' structure~\cite{petr} found by Horava~\cite{horava} 
to five dimensions.  
Hence, from the point of view of the dynamical fields in the bulk,  
the condensate on the wall looks like a $\delta$-function source in 
the equations of motion for the $Z_2$-odd  
hyperplet scalars. From the point of view of  
the Lagrangian, the {\em effective} coupling which corresponds to this 
interpretation is obviously  
\beq 
L_{coupling} = -\frac{1}{2}g_{xy} g^{55} (\pa_5 \sigma^x - \lc \delta (x^5 -  
\pi \rho) \delta^{x x_0}) (\pa_5 \sigma^y - \lc \delta (x^5 -  
\pi \rho) \delta^{y x_0}) 
\eeq 
where we assume, as mentioned above, the conventional  
wisdom that the four-dimensional gaugino condensate  
must be proportional to the Calabi-Yau $(3,0)$ form $\Omega_{ijk}$. 
and hence should couple to the $\sigma^{x_0}$ belonging to the 
universal hypermultiplet.  
 
Two types of singular sources will be  
present in the resulting equations of motion, 
namely $\delta$ functions and derivatives of these 
with respect to $x^5$. 
As the equations of motion are second-order differential equations,  
$\delta$-function singularities in them tell us that the derivatives 
of  
certain functions are discontinuous across the walls, and the presence of  
$\pa_5 \delta$ is the sign  
that a certain function is itself discontinuous across the wall.  
Quick inspection of the equations reveals that $\pa_5 \delta$-type  
source appear in the equations of motion for the odd scalar in 
the universal hypermultiplet. 
We infer that this scalar suffers a  
discontinuity across the wall of the form 
$\lc \theta (x^5 - \pi \rho)/2$, where $\theta(x)$ is  the Heaviside 
step function (since it is not $Z_2$-even, the limiting values 
on both  
sides must have the same magnitude, but opposite signs. On the other hand, 
$\pa_5C$ is even and continuous, so either the coefficient of the 
$\delta$-function singularity 
in the equation of motion for $C$ must vanish, or the explicit $\delta$ 
function must be cancelled by the implicit $\delta$ function developed by 
the  
solution of the equations along the direction of some other field or its  
derivative.  
Similarly, for the even moduli we conclude that either the coefficients of  
$\pa_5 \delta$-type operators have to disappear, or they are  
dynamically cancelled, as these fields can only have 
their derivatives with respect to $x^5$ 
discontinuous on the walls.  
The singularities are, for a generic scalar metric, distributed among all  
equations of motion, and mutual cancellation 
of all explicit and implicit singular operators is a good consistency check  
for the solution. It also helps to convert the system of equations  
with singular sources into the equivalent set of equations 
defined over the half-circle only, 
with suitable boundary conditions at the end of that half-circle. Details 
of this procedure shall be discussed using specific examples in Section 5.   
 
We now explore the forms of supersymmetry breaking 
admitted by the bulk theory at the ends of the $S_1 / Z_2$ line 
segment. There are important restrictions imposed by the  
$Z_2$ parity properties of the fields in the effective Lagrangian, 
since only $Z_2$-even objects may have non-zero vacuum 
expectation values on the walls. The fact that the $Z_2$ parity of any 
field $X$ is reversed by 
taking its derivative in the fifth direction $\pa_5 X$ 
increases the range of possibilities. It is clear that 
any candidate vacuum expectation values 
must also exhibit four-dimensional Lorentz invariance, which 
remains true for the $\pa_5$ derivatives.  
 
The five-dimensional Lagrangian (\ref{full5}) has the  interesting 
property that 
all the couplings, except those arising from the Riemannian connection, 
contain derivatives with respect to five space-time coordinates. 
Also, in the supersymmetry transformations  
(\ref{gtr}),(\ref{trvec}),(\ref{trhyp}), 
the terms on the right-hand sides are proportional to space-time 
derivatives 
or are multilinear in fermionic fields. We discard from our 
discussion the fermionic terms  
in supersymmetry transformations, on the grounds that 
their vacuum expectation values would be interpreted as the formation of  
bound states, which is not plausible, because the only  
force acting in the bulk is gravity.  Terms interesting for  
supersymmetry breaking must be able to survive the projection  
from the bulk onto the boundary walls. This means that we  
must consider the supersymmetry transformations of even combinations of 
fermions generated 
by an even combination of susy generators. 
Equipped with these observations, we now examine 
the supersymmetry transformation  
laws (\ref{gtr}),(\ref{trvec}),(\ref{trhyp}), for 
fermions in 
the different bulk supermultiplets, to see which of 
them may develop vacuum expectation values that might signal 
spontaneous supersymmetry breaking on the wall. 
 
In the case of the {\bf gravity supermultiplet}, 
one must consider separately the $\psi^i_5$ 
components of the gravitino. This is because, from the four-dimensional  
point of view which we  
have to assume when discussing eventually the supersymmetry breaking on  
the walls, they form a symplectic matter fermion, whose even  
component can in principle participate in supersymmetry breaking. 
The supersymmetry transformation law for the fifth component of the 
gravitino doublet is 
\beq 
\delta \psi_{5}^{ i}  =  ({\cal D}_5 \epsilon)^i - \omega_{xj}^i 
( \delta \sigma^x ) \psi^j_5 + \frac{i}{4 \sqrt{6}} h_I (\Gamma_5^{ 
\alpha \beta} - 4 \delta^\alpha_5 \Gamma^\beta) \epsilon^i F^I_{\alpha 
\beta} 
\label{gt5} 
\eeq 
As we argued earlier, we disregard fermionic terms in this transformation law, 
and concentrate on the covariant derivative. It contains a part  
originating from 
the Riemannian connection, and one related to the $Sp(1)$ connection: 
\beqa 
\delta \psi_{5}^{ i} & = & ({D}_5(\omega) \epsilon)^i - \omega_{xj}^i 
( \pa_5 \sigma^x ) \epsilon^j + \ldots  
\label{g5tr} 
\eeqa 
where the dots stand for the fermionic terms, and for terms containing  
Maxwell field strengths. We first consider the first term in (\ref{g5tr}), 
and decompose it as 
\beq 
{\cal D}_5 \epsilon^i = \pa_5 \epsilon^i + {1 \over 2} \omega_5^{ab} 
\Sigma_{ab} \epsilon^i : \; \Sigma_{ab} = {1 \over 4} [ \Gamma_a, 
\Gamma_b] 
\label{decompose} 
\eeq 
For the combination of $\epsilon^i$ which is 
even under $Z_2$, the 
first term in (\ref{decompose}) is odd, and hence cannot 
contribute. Expanding 
\beq 
\omega_5^{ab} = {1 \over 2} e^{\nu 5} (\pa_5 e^b_\nu - \pa_\nu e^b_5) 
- {1 \over 2} e^{\nu b} (\pa_5 e^a_\nu - \pa_\nu e^a_5) 
- e^{\rho a} e^{\sigma b} (\pa_\rho e_{\sigma c} - \pa_\sigma e_{\rho c}) 
e^c_5 
\label{omega} 
\eeq 
we see that the only possible non-vanishing vacuum expectation values 
would be those appearing in $\omega_5^{55}$, but this vanishes overall 
by antisymmetry. Turning now to the second term in (\ref{g5tr}), we note 
that the important connection here is that associated with the  
vielbeins $f^{ib}_x$, and that the exact form of $\omega_{xj}^i$ 
depends on the geometry of the scalar quaternionic manifold in question.  
For the simple model analyzed above,   
the explicit forms of the coefficients can be read off from the  
expressions given in Section 3, and for the left-handed and $Z_2$-even 
combination of $\psi_{5}^{i}$ one obtains 
\beqa 
\delta_{+} \psi_{5L}^{+}&=&\frac{\sqrt{2}}{\sqrt{S + \bar{S}} \sqrt{1-|Z|^2}} 
 (\pa_5 C_0 + Z \pa_5 C_1 ) \epsilon^{+}_{L}  
\label{psif} 
\eeqa 
which need not vanish on the wall. We conclude that the 
graviton hyperplet may in principle communicate supersymmetry breaking 
between the walls. 
 
Now, anticipating the need for the four-dimensional interpretation  
of five-dimensional results, we would like to identify the fields  
which survive the $Z_2$  projection in terms of 4d multiplets they are going to belong to.  
{}From the supersymmetry transformation $\delta e^m_5 = \frac{1}{2}  
\bar{\epsilon}^i \Gamma^m  \psi_{5 i}$, we see that  
the four-dimensional superpartner of the even part of  
$\psi_{5}^{ i}$ is $e^5_5$  the physical radius of the 5th dimension. 
The second scalar partner which completes the four-dimensional chiral  
$R_5$ supermultiplet is the   
$A_5$ component of the graviphoton. This can be seen, for instance, 
from the supersymmetry transformation of the 5th component of the  
graviphoton: 
$\delta A_5  =  \frac{i}{4} \sqrt{6}  \bar{\psi}^i_5 \epsilon_i  
$.  
This means, that the scalar superfield descending on the wall from the 5d  
gravity supermultiplet corresponds to the overal $T$ modulus identified  
within the simple truncation of the 5d supergravity to four dimensions 
in e.g.~\cite{dudas}. The projections of the remaining $h_{(1,1)}-1$   
multiplets containing vector fields shall correspond to the shape  
moduli supermultiplets 
called $T^i$, $i=1,\ldots, h_{(1,1)}-1\,$.    
 
To determine the possible r\^ole of these {\bf vector supermultiplets}, 
let us inspect the real metric moduli and vector fields.   
As one can easily  
check, all of the metric moduli, both real and complex, 
are even under $Z_2$. Hence, their   
derivatives with respect to $x^5$ do not exist on the walls, and  
they cannot be messengers  
of supersymmetry breaking between the walls. As for gauge bosons, the 
$A^I_{5}$ are also even, but the components $A^{I}_{\hat{\mu}}$ 
are odd. Hence, the derivatives $\pa_5 A^{I}_{\hat{\mu}}$  could  
in principle be  
useful. However, the gauge fields enter the Lagrangian  
through their field strengths 
only, and the corresponding components $F^I_{5\hat{\mu}}$ of all  
gauge field  
strengths have to vanish in the vacuum in order not to break 
four-dimensional Lorentz invariance. 
Specifically, the supersymmetric variations (\ref{trvec}) of the fermionic 
components of the vector supermultiplet are  
\beqa 
\delta \lambda^a_i & = & - i f^a_A (\not\!\pa t^A) \epsilon_i -  
\omega_A^{ab} (\delta t^A) \lambda^b_i + \frac{1}{4} h^a_I  
\Gamma^{\mu \nu} F^I_{\mu \nu} \epsilon_i  
\eeqa 
which after suppressing terms which must vanish because of Lorentz invariance 
gives 
\beq 
\delta \lambda^a_i = - i f^a_A (\Gamma^5\pa_5 t^A) \epsilon_i 
\label{t5vec} 
\eeq 
Since $t^A$ is 
even, $\pa_5 t^A$ is odd and hence must vanish on the walls. 
Hence, at leading order in the Calabi-Yau 
compactification of the $M$-theory Lagrangian, the vector 
supermultiplets cannot participate in the transmission 
of supersymmetry breaking between the walls~\footnote{However, 
it is possible that they  
may do so indirectly 
through mixing with the hypermultiplets in  
the equations 
of motion.}. 
 
Finally, we turn to the scalar {\bf matter hypermultiplets}. 
We have argued that the metric moduli, both real and complex, are  
$Z_2$-even. However, there exist complex scalars  
arising from the reduction of the  
internal part of the antisymmetric tensor, $C_{ijk},\,C_{ij\bar{k}}$,  
which are $Z_2$-odd, and hence have $\pa_5$ derivatives  
that are $Z_2$ even.  
These derivatives are the fields  that participate in the 
transmission of the supersymmetry breaking at the lowest order, and they 
form the $Z_2$-even 
bosonic components of the $h_{(2,1)}+1$ $N=2$ hypermultiplets.  
Consequently, it is a combination of the even fermions from these 
hypermultiplets which becomes the goldstino eaten up by the  
four-dimensional  
gravitino living on the three-brane to form a massive spin-$\frac{3}{2}$  
particle. Among the hypermultiplets there is one which is singled out,  
namely the so-called  
universal hypermultiplet. It is always present in Calabi-Yau-type  
compactifications, and it is associated with the  
unique $(3,0)$ form $\Omega_{ijk}$.    
The remaining hypermultiplets may be present or not, depending  
on the particular Calabi-Yau manifold chosen for compactification.  
However, they cannot be ignored if they are present,  
since, as  
discussed  
in detail in the next section, the $\sigma$-model metric in the  
hypermultiplet sector is usually highly non-trivial, and  
the equations of motion for the hypermultiplet  
scalars in the bulk are consequently non-linearly coupled, mixing together 
all of them, both even and odd, universal and non-universal. 
 
We now consider the supersymmetry transformation law (\ref{trhyp}) 
for the fermions in the scalar hypermultiplets ({\it `hyperinos'}): 
\beq 
\delta \lambda^b = - i f^{ib}_x ( \Gamma^5\pa_5 \sigma^x) \epsilon_i 
\label{t5hyp} 
\eeq 
where we have again suppressed terms that cannot contribute 
because of Lorentz invariance. Since $\pa_5 \sigma^x$ is even, 
the first term in (\ref{t5hyp}) may in principle contribute, 
if the form of the $f^{ib}_x$ is suitable. To see this 
more explicitly, 
we consider a single hypermultiplet, and write down   
the exact expression for the even supersymmetry transformation of an  
even fermion from such a multiplet~\footnote{The generalization of  
this formula to the case of more hypermultiplets is obvious.} 
\beqa 
\delta_{+} (\lambda^1_L+ i \lambda^2_L) & = & - \frac{i}{2} 
( f_{x1}^1 + f_{x2}^2 + i f_{x1}^2 -i f_{x2}^1 ) \pa_5  
\sigma^x \epsilon^{+}_L 
\label{crux} 
\eeqa 
where the summation over $x$ is over odd scalar components of the  
hyperplet. 
Now, let us call the complex scalars belonging to the universal  
hypermultiplet $(S,C)$ (the first is even, the second is odd), 
and the complex scalars belonging to the non-universal hyperlets  
$(Z^k, C^k)$, $k=1,\ldots,h_{(2,1)}$. The even scalars $Z^k$ are  
exactly the type $(2,1)$ complex metric moduli. As discussed 
in~\cite{subhar}, once the metric for these metric moduli is known, 
and using the observation that the full scalar-field space  
is a quaternionic manifold, one can reconstruct the full metric  
including the odd-field sector. The task is usually very difficult, 
but there exist simple examples, one of which we shall discuss in detail 
in the next section. From the point of view of four-dimensional  
boundaries, after performing the $Z_2$-projection,  
the universal hypermultiplet shall correspond to the chiral  
superfield denoted by $S$ in \cite{dudas}. This supermultiplet 
couples to the gauge kinetic terms on the walls, thus its scalar  
component which is the volume of the Calabi-Yau space, 
sets the tree-level gauge coupling if one ignores deformations.  
At the level of $\kappa^{4/3}$ deformations the coefficient of the  
gauge kinetic terms contains also the scalar part of the 4d field $T$  
which comes down from the 5d gravity supermultiplet, as we have noticed  
earlier in this section.  
 
To convince oneself that interesting  
supersymmetry breaking can indeed take place in the hyperplet sector,  
it is enough to consider the metric near one of the  
three-branes, i.e., up against the wall. Further, one can expand the 
metric around vanishing values  
of the $Z$-moduli, and explore order by order whether 
supersymmetry breaking happens. In fact, it is  
sufficient to inspect the  
zeroth-order case, in which the metric simplifies to  
\beq 
ds^2 = - {\cal K}^2 dZ d \bar{Z} - \frac{1}{(S + \bar{S})^2} 
dS d \bar{S} - \frac{2}{S + \bar{S}} d {\cal U}^T d \bar{{\cal U}} 
\eeq  
where ${\cal U}=(C,C^1,\ldots,C^{h_{(2,1)}})$.  
In this case, one can immediately solve for vielbeins with respect to the 
tangent-space metric $\Omega_{ib,jc}= - \frac{1}{2} \epsilon_{ij} 
\Omega_{bc}$, where $\Omega_{bc}$ is the constant antisymmetric  
matrix defining the $Sp(h_{(2,1)})$ group. Then, using the prescription  
(\ref{crux}), after some manipulations one finds that the variations  
of even fermions 
are of the form 
\beq 
\delta_{+} \lambda^{+} =   
\frac{1}{\sqrt{S + \bar{S}}} \, {\it \tilde{f}}_{k} (\pa_5 C)^{k}  
\label{beef} 
\eeq 
where $k=0,\ldots,h_{2,1}$, the `effective' vielbeins (numbered here by  
the `effective' index $k$) ${\it \tilde{f}}_{k}$ are numbers of order  
one plus corrections $o(Z)$ and the  
boundary  
values of $(\pa_5 C)^{k}$ are at this point arbitrary constants, 
which have no reason to be set to zero.  
This conclusion holds also in the limit where $h_{(2,1)}=0$, i.e., for the 
universal hypermultiplet alone. Let us note that  
$\frac{1}{\sqrt{S + \bar{S}}} \sim \frac{1}{\sqrt{ V_{CY}}}$  
where $V_{CY}$ is the  
Calabi-Yau volume,  
implying that the magnitude of supersymmetry breaking is directly 
sensitive to the scale of Calabi-Yau compactification.\\ 
 
We therefore see that - in contrast to the vector superplets - both the 
universal and non-universal hypermultiplets may 
contribute to supersymmetry breaking. This possibility is very 
interesting, as it indicates the possibility of non-trivial dynamics 
for the complex structure moduli, which have been largely ignored 
in studies of Calabi-Yau compactification of the the heterotic 
string in the weak-coupling limit. We have found that they may 
play a non-trivial r\^ole in the strong-coupling limit, 
represented by a non-trivial dependence on the additional 
coordinate. 
 
Finally, let us comment on the scenario where supersymmetry breaking occurs  
at some large distance scale $> R_5$.  
This possibility is similar to the `historical' scenario of 
supersymmetry breaking in the hidden sector discussed extensively 
in the weak-coupling limit of the heterotic string, where the length 
of the $S_1 / Z_2$ line segment is negligibly small. 
The main point we want to make here 
is that the procedure developed in this  
paper should also be used in this case, with the hard fermionic bilinear 
replaced  
by the effective superpotential for even moduli~\cite{laltom}.  
This would lead eventually to 
the possibility of a dynamical determination of the magnitude of the 
fifth dimension, but such an application goes beyond the scope of the  
present paper. 
 
\section{Solving the Equations of Motion} 
 
We now solve the five-dimensional  
equations of motion, with the goal of 
determining how information about physics from the hidden sector located 
on the hidden wall is fed into the observable sector of the theory.  
To obtain an answer to this question, one has to define the problem in a  
more precise way. One possible formulation would be to define 
five-dimensional observables  
accessible to the observer imprisoned on the visible wall. 
The other possibility is to explore the 
Kaluza-Klein paradigm, and seek an effective four-dimensional model, 
with the degrees of 
freedom corresponding to motions along the fifth dimension integrated out, 
which would summarize from the point of view of the truly four-dimensional  
observer 
the contribution of the physics on the hidden wall modulated by the  
dynamics in the bulk.  
For the time being, we explore the latter point of view. 
 
In view of the complicated nature of the generic field theory 
in the bulk, and in the absence of a well-defined 
non-perturbative model on the hidden wall, we study initially 
a toy scalar field model without excited gravitaional degrees 
of freedom, with minimal 
kinetic terms in the bulk and the simplest possible couplings to 
four-dimensional operators living on  
the walls. This helps us to clarify which is the proper way to compute  
the terms that may break supersymmetry: whether they are truly local  
operators on the visible wall, or whether one has to compute some 
non-local  
averages over the bulk. We conclude that the first of these two proposals is  
appropriate. 
 
We start with a scalar field that is even under $Z_2$, which has  
been discussed in ten dimensions in~\cite{low10,llo}.  
Let the action of the 
model be (from here on we shall use $\rho$ to denote the radius  
of the fifth dimension) 
\beqa 
S(\Phi) & = & \int d^5 x \left ( \frac{1}{2} \pa \Phi \pa \Phi - 
 \Phi {\cal S} \delta(x^5-\pi \rho) - {\cal O} \Phi \delta(x^5) \right ) 
\eeqa 
Here ${\cal O}$ is an operator composed of observable fields 
living on the visible wall at $x^5=0$, where we assume $<{\cal O}>_{vac} =0$ 
and $ \cs(x)$ is the function of the  
four-dimensional coordinates which  
represents hidden-wall sources, such as a hiddden gaugino condensate, 
which affect the vacuum configuration of the field. As pointed out 
in~\cite{low10}, the standard Kaluza-Klein procedure is inadequate here, as 
one  
encounters non-zero modes in the bulk which contribute to the 
four-dimensional equations of  
motion. In our case, the bulk equation of motion is  
\beq 
- \Box_{4} \Phi + \pa_5^{2} \Phi = \cs (x) \delta(x^5 -\pi \rho) + 
\co (x) \delta(x^5) 
\eeq 
The first step is to separate the zero mode, i.e., the part of the field  
which does not depend on the fifth coordinate. The decomposition should be  
of the form 
\beq 
\Phi = \phi(x) + \psi (x;x^5), 
\eeq 
but one must formulate a specific prescription for extracting the  
zero-mode field from the full $\Phi$ field. The natural and consistent 
prescription  
is the requirement that the average over the fifth dimension of the 
non-zero mode $\psi$ be zero, namely 
\beq 
\frac{1}{\pi \rho}  
\int_0^{\pi \rho} d x^5 
\psi(x;x^5) =0. 
\label{average} 
\eeq 
Next we point out that, in the present case of a 
$Z_2$-even field, 
all the information is contained on the 
half-circle between 0 and $\pi \rho$, 
once we apply the correct boundary conditions which, upon using the 
definite $Z_2$ parity, reproduce the singularity structure found 
in the equation of motion defined on the full circle~\footnote{Analogous  
reasoning also applies to $Z_2$-odd fields.}. 
In the case at hand,  
the boundary conditions are  
\beqa  
{}&{} & \lim_{x^5 \rightarrow \pi \rho} \pa_5 \Phi  = 
-\frac{\cs}{2} 
\nonumber \\  
{}&{} & \lim_{x^5 \rightarrow 0} \pa_5 \Phi  = 
 \frac{\co}{2} 
\label{bc1} 
\eeqa 
After the above-mentioned splitting, the equation of motion is  
\beq 
-\Box_4 \phi - \Box_4 \psi + \pa_5^{2} \psi = 0 
\eeq 
with the boundary conditions (\ref{bc1}). 
 
To proceed, we assume `low-energy' 
sources, whose derivatives along the space directions are not larger than 
the  
typical derivatives along the fifth dimension. Thus one can make for 
$\psi$ the series Ansatz of the form 
\beq 
\psi = \sum_{n=0}^{\infty} \psi_n 
\eeq 
with $\Box_4 \psi_n / \pa_5^{2} \psi_n \; < \; 1$, 
and write the equation of motion in the form 
\beq 
 -\Box_4 \phi -  \sum_{n=0}^{\infty}   \Box_4 \psi_n +  
 \sum_{n=0}^{\infty} \pa_5^{2} \psi_n = 0 
\eeq 
Boundary conditions on  
$\Phi$ now become boundary conditions on $\psi$. Solving in an obvious way  
the resulting series of equations, we obtain  
\beq 
\psi_0 = \Box_4 \phi ( \frac{(x^5)^2}{2} -\frac{\pi \rho^2}{6}) + \frac{\co}{2}  
(x^5 -\frac{\pi \rho}{2}) 
\eeq 
plus, from the boundary conditions, the relation $ \Box_4 \phi = -\frac{ 
\cs + \co }{2 \pi \rho}$. The remaining terms in the series can easily be computed  
from the recurrence equations $ -\Box_4 \psi_{n-1} + \pa_5^{2} \psi_n = 0 
$  
equipped with trivial boundary conditions. When we substitute the solution  
$\Phi = \phi + \psi_0$ into the action, and perform the integration over the  
circle in the fifth dimension, we obtain the effective action 
\beqa 
S_{eff} ( \phi )& =  & \int d^4 x \left ( \pi \rho \pa_{\mu} \phi \pa^{\mu} \phi  
+ \frac{\pi \rho^3}{180} (\pa_{\mu} \cs  \pa^{\mu} \cs + \pa_{\mu} \co  \pa^{\mu} \co 
- \frac{7}{4} \pa_{\mu} \co \pa^{\mu} \cs ) \right ) \nonumber \\ 
&-& \int d^4 x ( \phi +\psi_0 )_{x^5=0} \co (x) - \int d^4 x ( \phi +\psi_0 )_{x^5=\pi \rho} \cs (x)  
\label{efev} 
\eeqa 
which, after substituting the values of the known solution 
for $\psi_0$ at the boundaries, becomes 
\beqa 
S_{eff} ( \phi )& =  & \int d^4 x \left ( \pi \rho \pa_{\mu} \phi \pa^{\mu} \phi  
- \phi (\cs + \co) 
+  \frac{\pi \rho^3}{180} (\pa_{\mu} \cs  \pa^{\mu} \cs + \pa_{\mu} \co  \pa^{\mu} \co 
- \frac{7}{4} \pa_{\mu} \co \pa^{\mu} \cs ) \right ) \nonumber \\ 
&+& \int d^4 x ( \co^2 + \cs^2 - \cs \co ) \frac{\pi \rho}{6} 
\label{efev2} 
\eeqa 
where the last line has arisen from the boundary terms in (\ref{efev}). 
 
The point to be noticed is that the effective operator that mixes sources  
on the hidden wall and observable fields on the visible wall is proportional  
to $\rho$, the distance between the walls. The message of this  
example applies also to  
other even fields living in the bulk, such as components of the gravity  
multiplet. The factor of $\rho$, which is present there  
also in the presence of boundaries, is in agreement 
with the prescription for obtaining the four-dimensional gravity  
action given in~\cite{strw}.  
 
We now consider a second case, which is of particular interest in the 
context of $M$ theory, namely 
a field which is odd under the action of $Z_2$.  
In this case, the Lagrangian which includes a coupling to sources on the 
hidden wall 
and to operators consisting of observable fields on the visible wall is 
\beqa 
S(\Phi) & = & \int d^5 x \left ( \frac{1}{2} \pa \Phi \pa \Phi + 
\pa_5 \Phi {\cal S} \delta(x^5-\pi \rho) + \pa_5 \Phi {\cal O} \delta(x^5) \right ) 
\eeqa 
The bulk equation of motion is  
\beq 
- \Box_4 \Phi + \pa_5^{2} \Phi = \cs (x) \pa_5 \delta(x^5 -\pi \rho) +  
\co (x) \pa_5 \delta(x^5) 
\eeq 
It is clear that the zero mode, i.e., the part of the field  
which does not depend on the fifth coordinate, must vanish for the odd 
fields.  
The equation of motion for the non-zero mode $ \psi (x;x^5)$ coincides then 
with the full equation of motion,  
the proper boundary conditions on the half-circle being  
\beqa  
{}&{} & \lim_{x^5 \rightarrow \pi \rho} \psi=  
\frac{1}{2} \cs   
\nonumber \\  
{}&{} & \lim_{x^5 \rightarrow 0} \psi =  
- \frac{1}{2} \co  
\eeqa 
Again, using the same method as for the even field,  
one finds the solution  
\beq 
\psi = \frac{\cs + \co}{2\pi \rho} x^5 - \frac{\co}{2} + \psi_1 + \ldots 
\label{e66} 
\eeq 
where the higher terms in the series can  
easily be computed from the recursive relation  
$ -\Box_4 \psi_{n-1} + \pa_5^{2} \psi_n =0$ with trivial  
boundary conditions. One should note that if the sources  
can be regarded as constants, independent of $x$,  
then the lowest-order solution indicated above  
is exact.  
 
After substituting the lowest-order solution into the  
action, one obtains the contribution from the odd field to the effective  
four-dimensional Lagrangian, including boundary terms: 
\beqa 
\delta S_{eff}&=& \int d^4 x \left ( \frac{\pa_{\mu} \cs   
\pa^{\mu} \cs }{12} \pi \rho + \frac{\pa_{\mu} \co   
\pa^{\mu} \co }{12} \pi \rho  
- \frac{\pa_{\mu} \cs   \pa^{\mu}  \co }{12} \pi \rho \right ) 
\nonumber \\ 
&-&  \int d^4 x ( \pa_5 \psi \co (x))_{x^5=0} 
-  \int d^4 x ( \pa_5 \psi \cs (x))_{x^5=\pi \rho} 
\label{oef} 
\eeqa 
With the help of (\ref{e66}) one computes the $\pa_5$-derivative of  
$\psi$~\footnote{This is immediate in the present case,  
but would demand solving the actual equations of motion in realistic 
models} 
\beq 
\pa_5 \psi = \frac{\cs + \co}{2\pi \rho} + \ldots   
\label{der66} 
\eeq 
The observation to be made here, which easily generalizes to  
more realistic models, is that the $\pa_5 \psi$, hence its  
values computed on the walls as well, is inversely proportional  
to the distance $\pi \rho$ between the walls.  
After substituting (\ref{der66}), applied on 
the walls, into (\ref{oef}) 
one finds the effective local Lagrangian in the form: 
\beq 
\delta S_{eff} = \int d^4 x \left ( \frac{\pa_{\mu} \cs   
\pa^{\mu} \cs }{12} \pi \rho + \frac{\pa_{\mu} \co   
\pa^{\mu} \co }{12} \pi \rho  
- \frac{\pa_{\mu} \cs   \pa^{\mu}  \co }{12} \pi \rho - \frac{3}{4 \pi \rho} 
 ( \cs \cs + \co \co + 2 \cs \co) 
\right ) 
\eeq 
As there is no zero mode, one obtains only an effective coupling of the source  
to the observable operators, and a contribution to the vacuum energy.  
 
There are two points to be made on the basis of the above arguments.  
First, the effective coupling of the source to the observable operators is  
born {\em locally}, on the visible wall, through the {\em local} value of  
the bulk variable, either the field itself or its derivative, 
as both 
come from terms which contain $\delta(x^5)$ coefficients. This is in  
perfect agreement with the fact that we are using only local solutions of  
the classical equations of motion to derive the effective action, and  
no quantum-mechanical effects are involved~\footnote{In other words, we 
use  
the saddle-point trajectory to compute the respective path integral, 
without  
including fluctuations even at the Gaussian level.}.   
Hence the whole influence of the bulk physics is encoded in the  
$x^5$ dependence of the classical trajectory, and the sources  
on the hidden wall simply play the r\^ole of boundary conditions for that  
classical trajectory. This feature becomes more significant in the case of 
more complicated 
non-linear $\sigma$ models in the bulk, when the leading dependence on 
$x^5$  
of the even and odd fields is less simple, with fields varying 
significantly over the distance between the walls.   
Secondly, one should note that in the case of the even field the strength  
of the effective $\cs \co$ interaction grows with the distance $\pi \rho$,  
whereas in the case of the odd field the strength decreases like  
$\frac{1}{\pi \rho}$.  
Finally, one should notice that, in more realistic models, the source 
$\cs(x)$ will be in fact a function of `boundary' configurations  
of the even bulk fields: $\cs (x) = \cs (E^i (x;\; x^5=\pi \rho))$, which 
will introduce additional aspects into the analysis. For the moment, 
we limit ourselves to remarking that, despite these complications, if the 
local dynamics 
on the hidden wall produces a stable configuration of the  
local Lagrangian truncated to that wall, the final  
conclusions for trajectories close to 
that wall-bound `vacuum' resemble those above.   
  
We now tackle the non-linear $\sigma$ model 
equations of motion in the bulk.  
We first warm up with the simple example where the scalars of 
non-universal hypermultiplets are turned off.  
As one can quickly compute using the quaternionic metric given in Section 3, 
the equations of motion for the coupled system of the even scalar $S$ and the  
odd universal scalar $C$ are (along the real directions) 
\beqa 
S''(x^5) &+& 
{{\delta(x^5 -\pi \rho)}^2}\,\left( {\frac{4\,{{\var}^2}\, 
          {{C(x^5)}^2}}{2\, 
           {{C(x^5)}^2} - S(x^5)}} -  
      {\frac{2\,{{\var}^2}\,S(x^5)} 
        {2\,{{C(x^5)}^2} - S(x^5)}} \right) -  
   {\frac{4\,{{C(x^5)}^2}\, 
       {{C'(x^5)}^2}}{2\, 
        {{C(x^5)}^2} - S(x^5)}} +  \nonumber \\ 
 & &   {\frac{2\,S(x^5)\,{{C'(x^5)}^2}} 
     {2\,{{C(x^5)}^2} - S(x^5)}} -  
   {\frac{4\,C(x^5)\,C'(x^5)\, 
       S'(x^5)}{2\,{{C(x^5)}^2} - S(x^5)}} +  
   {\frac{{{S'(x^5)}^2}} 
     {2\,{{C(x^5)}^2} - S(x^5)}} = 0 \nonumber \\ 
C''(x^5)& +& 
  {\frac{4\,{{\var}^2}\,C(x^5)\, 
       {{\delta(x^5-\pi \rho)}^2}}{2\,{{C(x^5)}^2} - S(x^5)}} 
     - {\frac{4\,C(x^5)\, 
       {{C'(x^5)}^2}}{2\, 
        {{C(x^5)}^2} - S(x^5)}} + \left( {\frac{-2\,\var\, 
          {{C(x^5)}^2}}{2\, 
           {{C(x^5)}^2} - S(x^5)}} \right . \nonumber \\ 
 &+&  \left .   {\frac{\var\,S(x^5)} 
        {2\,{{C(x^5)}^2} - S(x^5)}} \right) \, 
    \delta'(x^5-\pi \rho) + {\frac{C'(x^5)\,S'(x^5)} 
     {2\,{{C(x^5)}^2} - S(x^5)}}  =  0   
\label{quat2} 
\eeqa 
where we use for convenience a stiff source 
$\var$, possibly representing a gaugino 
condensate at short distances, which replaces the general  
source $\cs$ from previous paragraphs. 
As they stand, the equations are defined on the full circle, 
corresponding to  
the fifth coordinate $x^5$ varying between $0$ and $2 \pi \rho$. However, we know 
that, because of the definite parities of the fields, all the information 
is contained  
on the half-circle $[0,\pi \rho]$. To write the equivalent problem on the 
half-circle, 
we need to define the proper boundary conditions which, together with the 
known  
parities, reproduce the singular sources located at the ends of the  
half-circle. As it is easy to see, the proper boundary conditions are 
\[ 
\lim_{x^5 \rightarrow 0} C=0, \; 
\lim_{x^5 \rightarrow \pi \rho} C=\frac{\var}{2} 
\] 
\begin{equation} 
\lim_{x^5 \rightarrow 0} S'=0, \; 
\lim_{x^5 \rightarrow \pi \rho} S'=0 
\label{conds} 
\end{equation} 
One can easily see that the resulting singularity structure on the full  
circle is the proper one: with the above assignment, $C'$ develops a  
singularity at $x^5 = \pi \rho$, $\lim_{x^5 \rightarrow 0} C= \frac{\var}{2} 
\theta (x^5-\pi \rho)$, and when one substitutes this into the equations (\ref{quat2}) 
all the singularities indeed cancel among themselves at $x^5 = \pi \rho$.  
We note that the solutions to these equations  
develop  singularities inside the perfect square describing the  
coupling of bulk to  
the boundary, which cancel exactly the singularity associated with  
the source. To see this cancellation in the equations on the full circle, 
it is essential to keep the $\delta^2$-type terms in the equation, despite the  
appearance that they are of higher order in the bulk-wall coupling.  
  
The structure of the equations (\ref{quat2}) is such that one can easily  
order the terms according to the powers of the moduli $C$, which can be 
taken as small as one wishes. The zeroth-order terms in such an expansion  
are second derivatives $S'', \; C''$ together with singular sources  
proportional to $S$, and to that order the second equation coincides with the  
simple equation studied in~\cite{peskin}.  
The system therefore approaches the toy 
models studied at the beginning of this Section.  
This explains why the qualitative  
behaviour of our solutions can be understood in terms of the toy models, and  
parallels the results obtained in~\cite{peskin}.    
 
At this point we would like to turn our attention to the possibility of  
including a  
non-trivial background of the even scalar $S$, whose real part corresponds 
to the volume of the internal Calabi-Yau space,  
which varies linearly with $x^5$  
according to the vacuum solution  
obtained by Witten~\cite{strw} in eleven dimensions. 
This can be achieved by adding sources on the right-hand side 
of the first of the equations (\ref{quat2}). The modified equation with the  
sources is  
\beqa 
S''(x^5) &+& 
{{\delta(x^5-\pi \rho)}^2}\,\left( {\frac{4\,{{\var}^2}\, 
          {{C(x^5)}^2}}{2\, 
           {{C(x^5)}^2} - S(x^5)}} -  
      {\frac{2\,{{\var}^2}\,S(x^5)} 
        {2\,{{C(x^5)}^2} - S(x^5)}} \right) -  
   {\frac{4\,{{C(x^5)}^2}\, 
       {{C'(x^5)}^2}}{2\, 
        {{C(x^5)}^2} - S(x^5)}} +  \nonumber \\ 
 & &   {\frac{2\,S(x^5)\,{{C'(x^5)}^2}} 
     {2\,{{C(x^5)}^2} - S(x^5)}} -  
   {\frac{4\,C(x^5)\,C'(x^5)\, 
       S'(x^5)}{2\,{{C(x^5)}^2} - S(x^5)}} + \nonumber \\ 
 & &  {\frac{{{S'(x^5)}^2}} 
     {2\,{{C(x^5)}^2} - S(x^5)}} = - \varrho_{v} \delta (x^5) +  
\varrho_{h} \delta (x^5 - \pi \rho) 
\label{ssrc} 
\eeqa 
and the corresponding boundary conditions on the half-circle,  
replacing (\ref{conds}), are 
\beq 
\lim_{x^5 \rightarrow 0} S'= -\frac{\varrho_{v}}{2},\; 
\lim_{x^5 \rightarrow \pi \rho} S'=-\frac{\varrho_{h}}{2}  
\label{ssrcb} 
\eeq  
where $\varrho_{v,h}$ are fixed parameters determined by solving 
the Bianchi identities. 
Again, one can check that the singularities cancel between themselves.  
One notices that the only consistent way to assign a value to the  
$Z_2$-odd function on the wall, even if the function has a finite 
discontinuity, is to give it the value zero at $x^5=0$ and $x^5=\pi \rho$.  
The results of solving the modified equations (\ref{quat2})  
in various cases are given in 
Figs.~1 to 9. In practice, the way solutions are obtained is to assume,  
among other boundary conditions, a value of $S(x^5 =0)$ and a value  
of $S'(x^5 =0)$. The former is unconstrained, the latter is interpreted  
in the spirit of (\ref{ssrcb}) as the presence of a source $\varrho_{v}$. 
Such boundary conditions generate typically a nonzero $S'(x^5 = \pi \rho)$. 
Again, this we simply interpret as a source $\varrho_{h}$ located at  
$x^5 = \pi \rho$. This approach is sufficient for our purposes, as we  
are interested in correlations between qualitative characteristics  
of the system displayed in the Figures, and not in obtaining  
any particular values of the sources. 
 
We use our numerical solutions of the equations of motion, 
with the above boundary conditions, 
to find the variations of the fields through the bulk, 
and to determine the boundary values of the interesting quantities on the  
visible and hidden walls. From the point of view of supersymmetry breaking,  
the interesting quantities are, as argued above and in Section 4, the  
local values of derivatives of the $Z_2$-odd fields on the walls,  
such as $\pa_5 C$ 
in the present case, modulated by the vielbeins. The vielbeins, as can be  
checked using expressions given in Section 4, are factors ${\cal O}(1 /  
\sqrt{2 V_{CY} (x^5)})$ times factors of order one. Here we display  
the leading dependence on $S$ (the $V_{CY}$) of the  
supersymmetric variations of hyperinos.   
The solutions presented below 
generically have supersymmetry broken both on the hidden and on the 
visible walls.  
\par 
\centerline{\hbox{ 
\psfig{figure=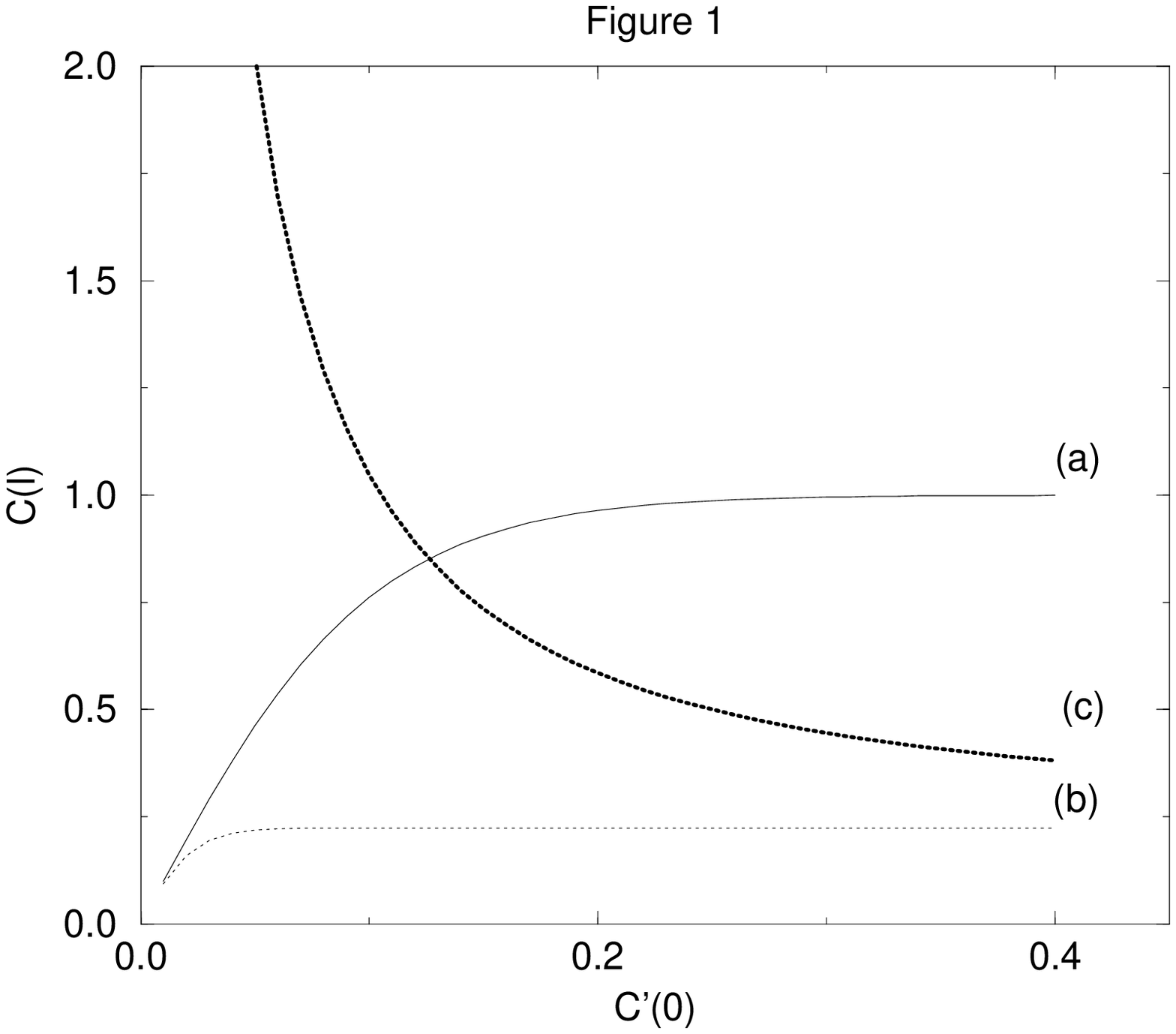,height=3.0in,width=3.0in} 
}} 
\par 
{\small Figure 1: {\em The relation between the value of $C'(x^5=0)$,  
measuring the supersymmetry breaking on the visible wall,  
and the limiting value of $C(x^5=\pi \rho)$, which equals half of the  
effective condensate on the hidden wall. Curves (a) and (b) 
correspond to $S'=0$ and $S=1,\;0.05$ respectively.  
Curve (c) is for $S=0.05$ and the non-zero slope  
$S'(0)=0.1$. These results are obtained for the model  
with only a universal hypermultiplet.}} \\ 
\vspace{0.3cm} 
 
We see in Fig. 1 the relation between the value of $C'(x^5=0)$,  
which measures the supersymmetry breaking on the visible wall,  
and the limiting value of $C(x^5=\pi \rho)$, which equals half of the  
effective condensate on the hidden wall. Curves (a) and (b) 
correspond to $S'=0$ and $S=1,\;0.05$ respectively, $\pi \rho =10$ is assumed in all  
examples and, as everywhere in this paper, we work in units where  
the eleven-dimensional Planck scale is equal to 1. The results shown in 
the 
figures are generic. One finds that a non-zero hidden condensate 
generically  
corresponds to non-zero supersymmetry breaking on the visible wall. Also,  
in the range where the variables are sufficiently small, 
corresponding to realistic supersymmetry breaking that is 
hierarchically smaller than the Planck scale, 
the scale of the visible supersymmetry breaking grows with the  
scale of the condensate. It is also clear, comparing curves (a) and (b)  
in the Fig. 1, that the change of $S(0)$ does not change the 
qualitative character 
of the relation between the quantities, just the numerical 
magnitude of the effect. The curves (c) tell us what happens  
when a non-trivial background along the direction of the field $S$  
is switched on, see (\ref{ssrc}) and (\ref{ssrcb}).  
This corresponds directly to the linear dependence  
of the volume of the Calabi-Yau space on the fifth (eleventh) coordinate  
found in~\cite{strw}, since $\Re (S) = V(x^5)$. In our model, this  
effect is taken into account by switching on the derivatives of $S$  
with respect to $x^5$ on the boundary of the semi-circle~\footnote{One 
should remember the 
parity properties when going over to the full circle and  
incorporating singular sources.}. 
We see that assuming $S'(0)$ non-zero makes the  
dependence of the visible supersymmetry breaking on the condensate  
become increasingly steeper. Also, this enhances the inhomogeneity of the  
field configurations in the bulk. Curve (c) is given for $S=0.05$ and  
the non-zero slope (quasi-linear $S$ background) $S'(0)=0.1$.      
To see how inhomogenous the bulk becomes, we plot the $S$ and $C'$  
fields as functions of $x^5$ in Figs. 2 and 3.  
\par 
\centerline{\hbox{ 
\psfig{figure=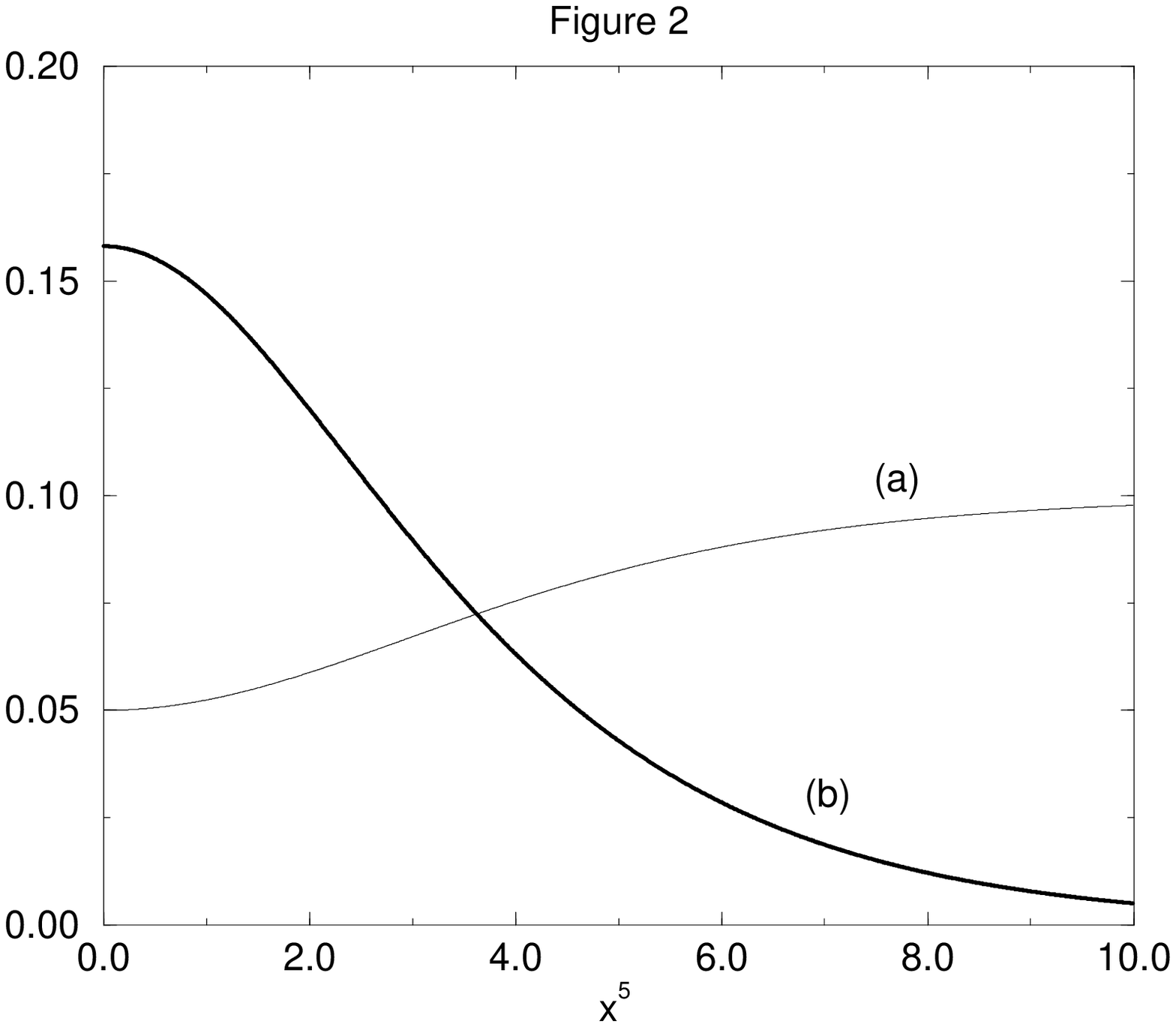,height=3.0in,width=3.0in} 
\psfig{figure=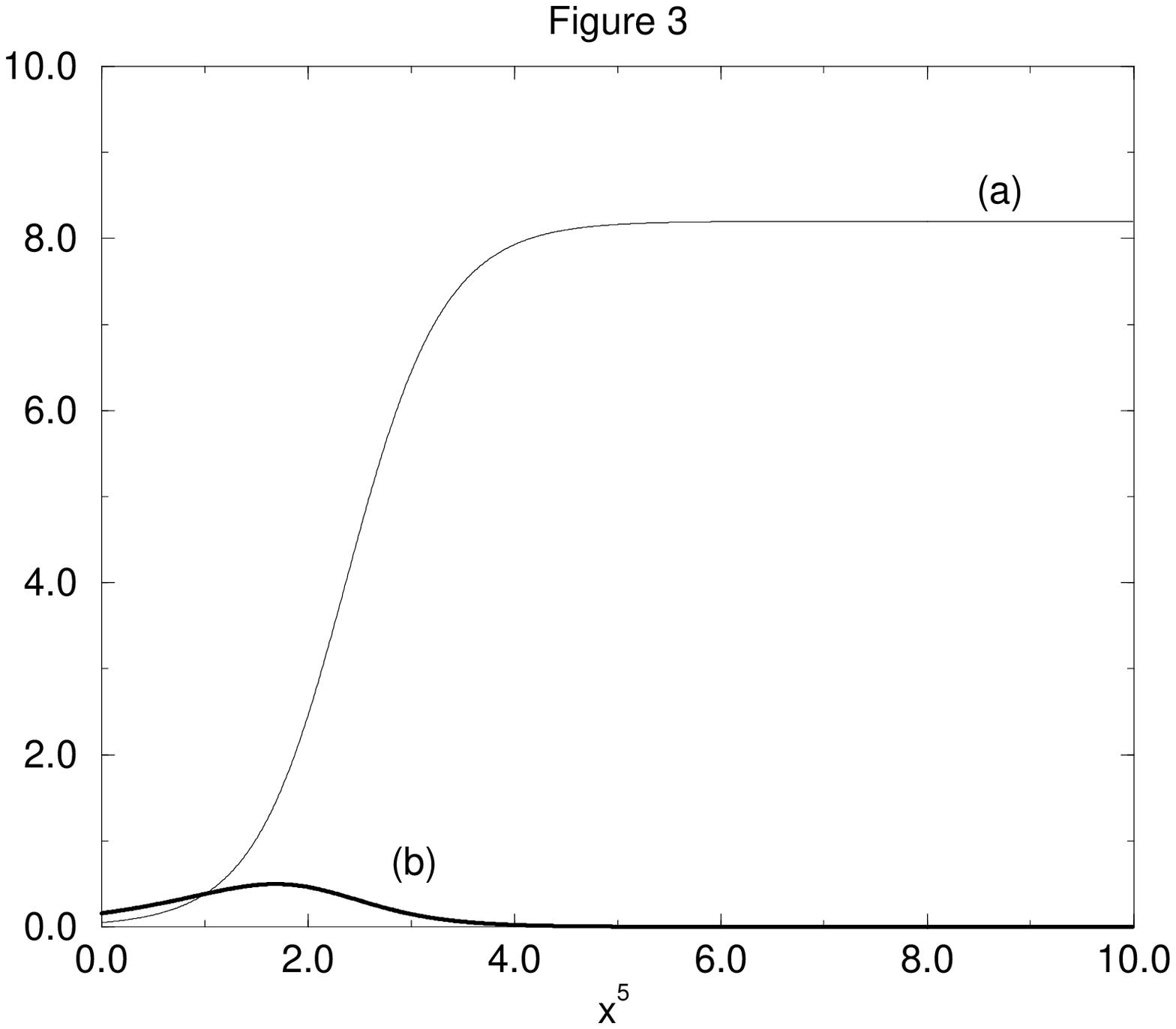,height=3.0in,width=3.0in} 
}} 
\par 
{\small Figure 2: {\em The $x^5$ dependence of the moduli $S(x^5)$ 
- curve (a) - and  
$\frac{C'(x^5)}{\sqrt{2 S(x^5)}}$ - curve (b). The boundary values of  
the fields on  
the visible wall $x^5=0$ are $S(0)=C'(0)=0.05$, $S'(0)=0$. 
These results are also for the model with just the universal 
hypermultiplet}.\\ 
 
Figure 3: {\em The same as in Figure 2, but with $S'(0)=0.1\;$.}} \\ 
\vspace{0.3cm} 
 
One can easily see that the field configurations  
are not homogeneous,  
and not even linear in $x^5$. The non-zero initial slope $S'(0)$ makes the  
variation of both plotted fields much more pronounced.  
Simple averaging over the fifth dimension does not commute with 
the non-linearity of the  
equations of motion in the bulk, and does not produce particularly natural  
characteristics of the behaviour of individual fields in the bulk.    
 
It is straightforward to extend the procedure described above to the case  
of the system containing one non-universal hyperplet, which includes one  
odd scalar $C_1$ and one even modulus $Z$,  
again making use of the quaternionic metric given in Section 4.  
The boundary conditions which faithfully  
reproduce the singularity structure in this case are 
\beqa 
\lim_{x^5 \rightarrow 0} C=0,\; 
\lim_{x^5 \rightarrow \pi \rho} C=\frac{\var}{2},&& \nonumber \\ 
\lim_{x^5 \rightarrow 0} C_1 =0, \; 
\lim_{x^5 \rightarrow \pi \rho} C_1 = 0,&& \nonumber \\ 
\lim_{x^5 \rightarrow 0} S'= -\frac{\varrho_{v}}{2} , \; 
\lim_{x^5 \rightarrow \pi \rho} S'= -\frac{\varrho_{h}}{2},&& \nonumber \\ 
\lim_{x^5 \rightarrow 0} Z'=0,\; 
\lim_{x^5 \rightarrow \pi \rho} Z'=0&& 
\eeqa  
where $\vartheta$ is the vacuum expectation value of the condensate.  
Again, one can solve the full set of equations using these boundary  
conditions. The results are presented below, and are qualitatively  
similar to the simpler case. The interesting new phenomenon  
is to watch is the r\^ole of the mixing between the two odd fields. It 
turns out that, even though the non-universal field $C_1$ is assumed not 
to couple directly to any source of supersymmetry breaking on the 
wall~\footnote{This assumption 
is not crucial, and there are models where it is relaxed.},  
the mixing with the remaining fields excites its derivative $\pa C_1$  
on the visible wall.  
\par 
\centerline{\hbox{ 
\psfig{figure=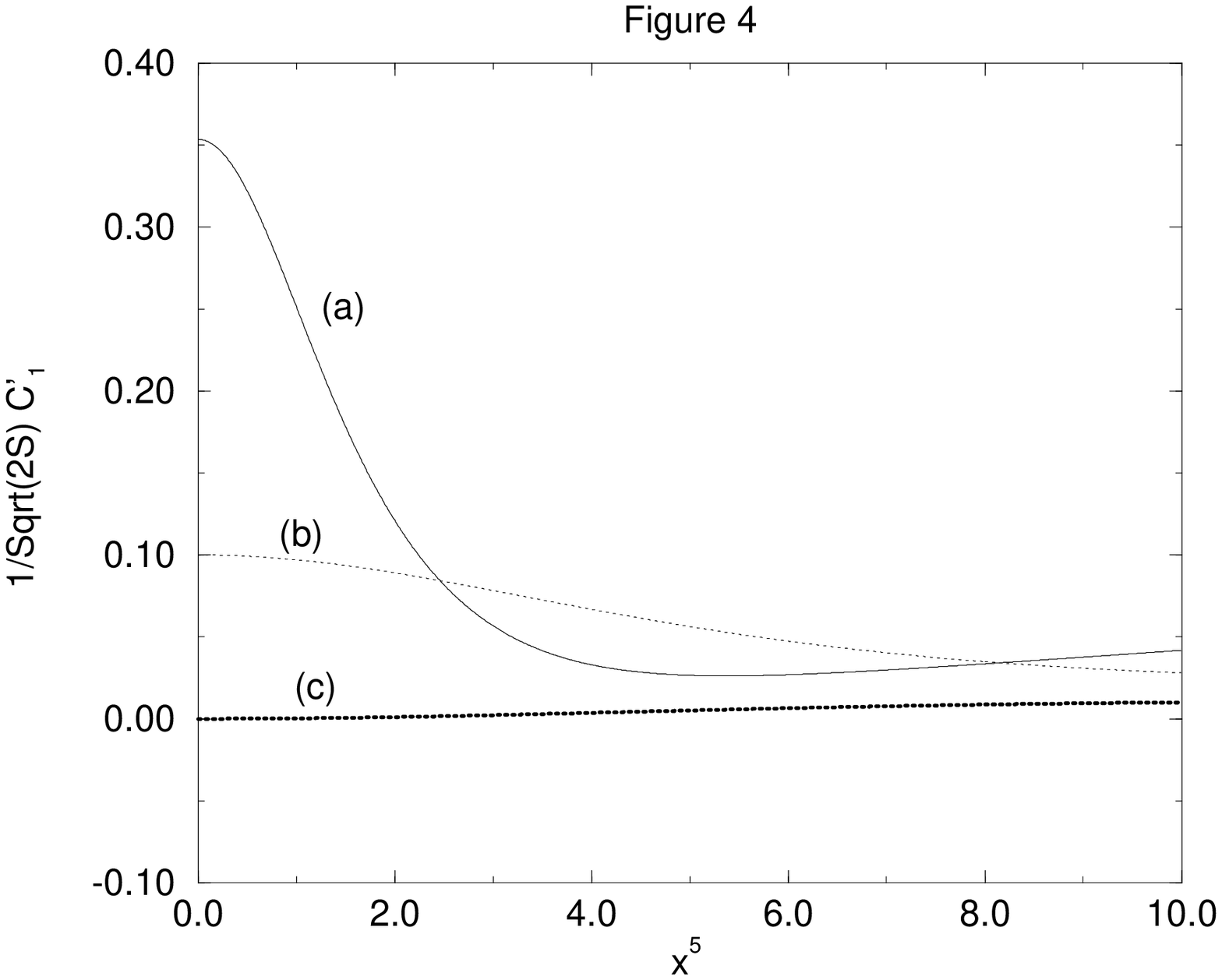,height=3.0in,width=3.0in} 
\psfig{figure=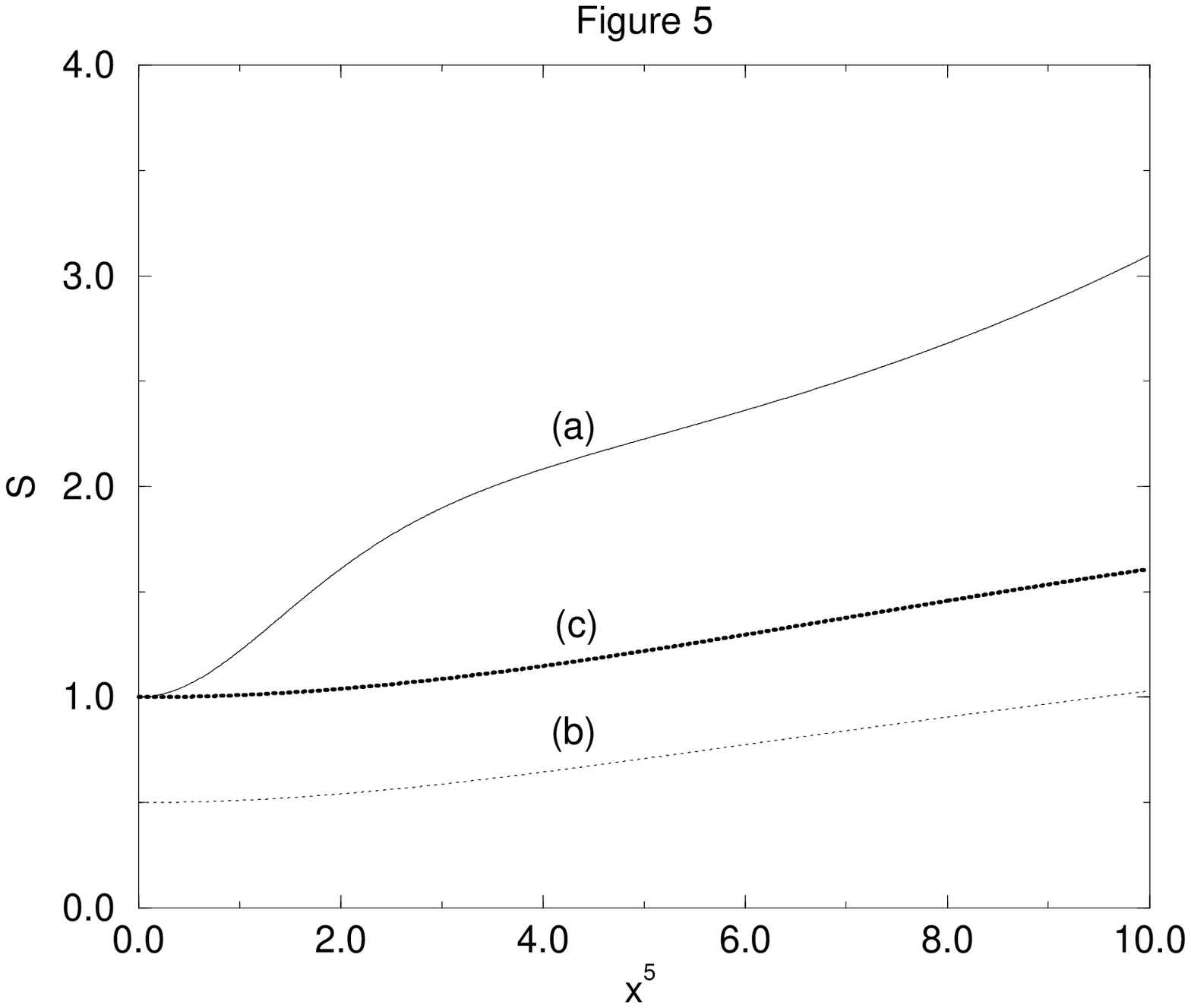,height=3.0in,width=3.0in} 
}} 
\par 
{\small Figure 4: {\em The $x^5$ dependence of   
$\frac{C_{1}'(x^5)}{\sqrt{2 S(x^5)}}$.  
Here $C_1'$ is the derivative with respect to  
$x^5$ of the $Z_2$-odd non-universal hyperplet modulus. Curves correspond 
to different boundary conditions on the visible wall. Curve (a) corresponds to  
$S(0)=1,\;Z(0)=0.1, \; C_{1}'(0)=0.5,\; C'(0)=0$,  
curve (b) to $S(0)=0.5,\;Z(0)=0.1, \; C_{1}'(0)=0.1,\; C'(0)=0.1$, and (c) 
to $S(0)=1,\;Z(0)=0.1, \; C_{1}'(0)=0,\; C'(0)=0.5$. 
Here and in the subsequent two Figures the model used contains a single 
non-universal hypermultiplet.}\\ 
 
Figure 5: {\em The evolution of the volume modulus $S$  
across the bulk. The boundary conditions for the curves are the same as in 
Figure 4.}} \\ 
\vspace{0.3cm} 
 
It is interesting to note that, even if one starts with vanishing  
derivative $C_{1}'$ 
at one boundary, as in curve (c) in Fig.~4, this derivative 
is excited  
in the bulk through the mixing with other moduli in the bulk Lagrangian.  
The field $S$ (volume modulus) changes visibly across the bulk,  
as seen in Fig.~5, signalling the  
necessity of introducing  sources on the second wall to maintain  
consistency between the boundary behaviour on the  
semicircle and the $Z_2$ parity properties.  
 
In the Figure 6 we have sketched the correlation between boundary value of  
the derivative $C_{1}'$ at the visible wall and the vacuum expectation 
value of the 
stiff condensate at the hidden wall, if the remaining boundary  
parameters are fixed. The visible 
correlation proves the influence of the  
$\sigma$-model dynamics on the evolution of the fields, as the 
modulus  $C_1$ does not couple directly to the condensate.  
\par 
\centerline{\hbox{ 
\psfig{figure=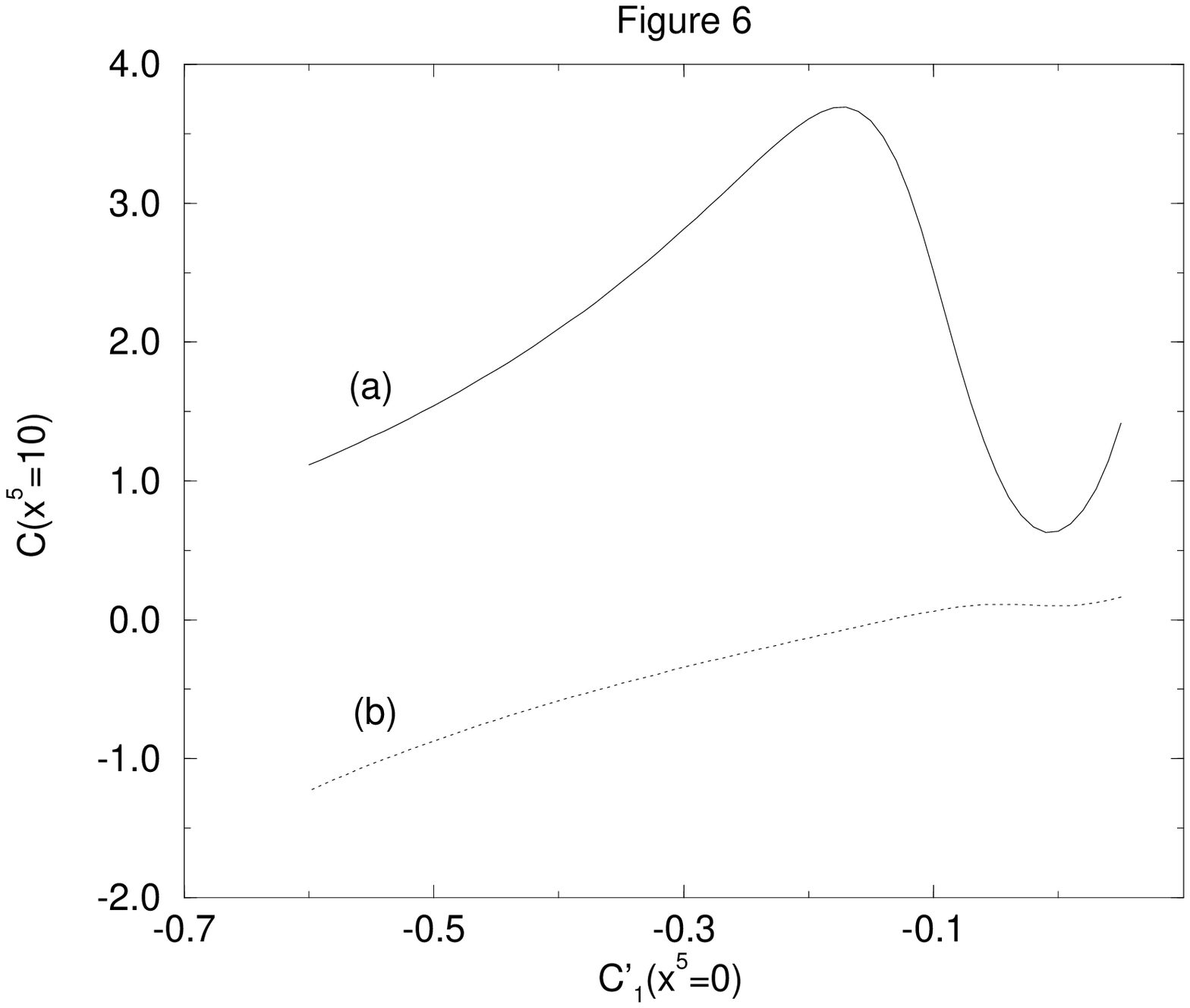,height=3.0in,width=3.0in} 
\psfig{figure=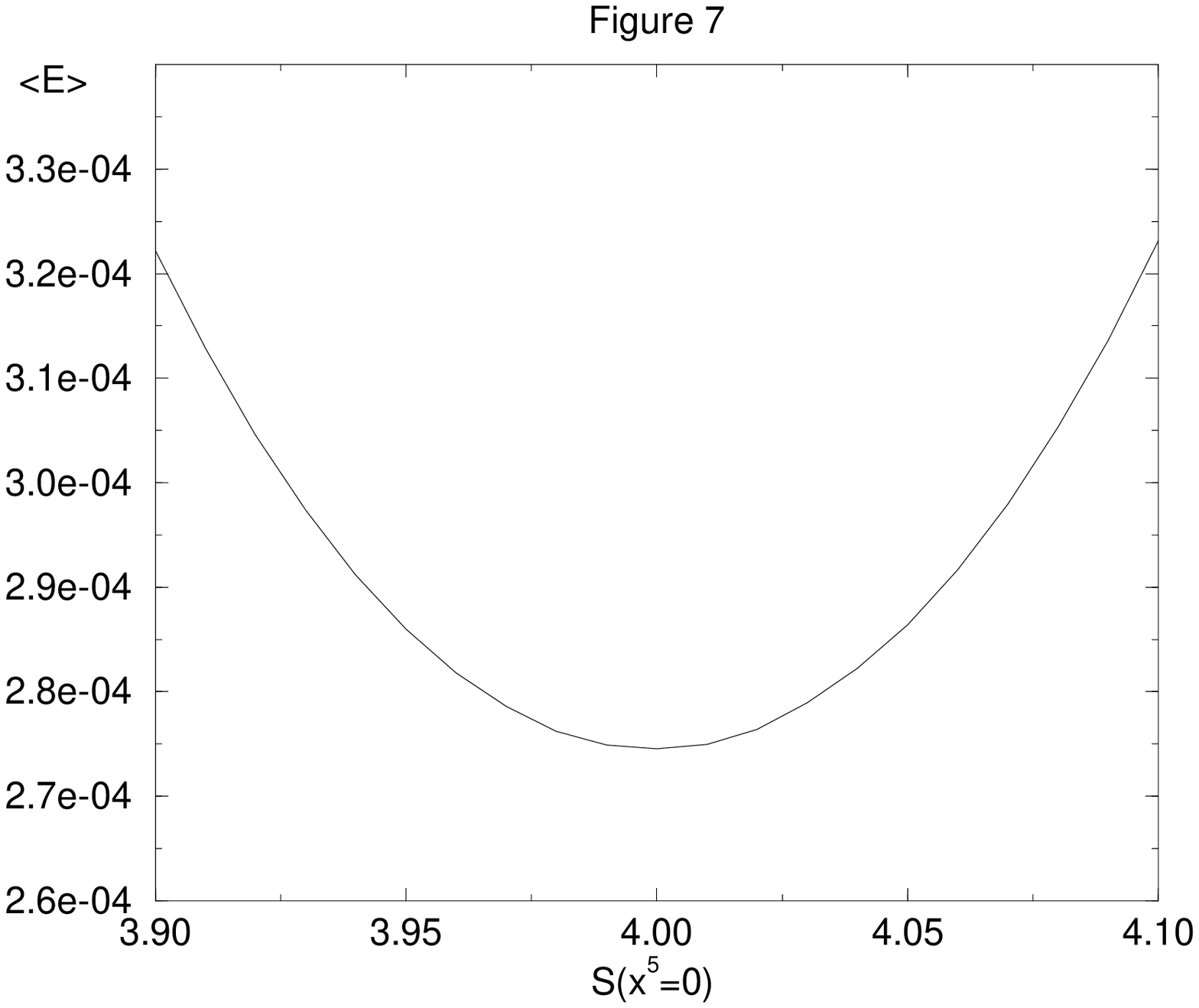,height=3.0in,width=3.0in} 
}} 
\par 
{\small Figure 6: {\em The correlation between boundary value  
of $C_{1}'$ at the visible wall and the vacuum expectation value  
of the stiff condensate at the hidden wall.  
Curve (a) corresponds to $S(0)=0.5,\;Z(0)=0.1,\; C'(0)=0.1$, and 
curve (b) differs in the boundary value $C'(0)=0.01$.}\\ 
 
Figure 7: {\em The vacuum expectation value of the Hamiltonian in the case of  
the two-piece boundary potential for the volume modulus $S$, as a 
function of the  
boundary value of $S$ at the visible wall. The local minimum of the part of the  
potential on the visible wall is $S_{0,vis}=4.0$ and on the hidden wall 
$S_{0,hid}=1.0$. Here and in the following two Figures the modified 
model (75, 76) is used.}} \\ 
\vspace{0.3cm} 
\par 
\centerline{\hbox{ 
\psfig{figure=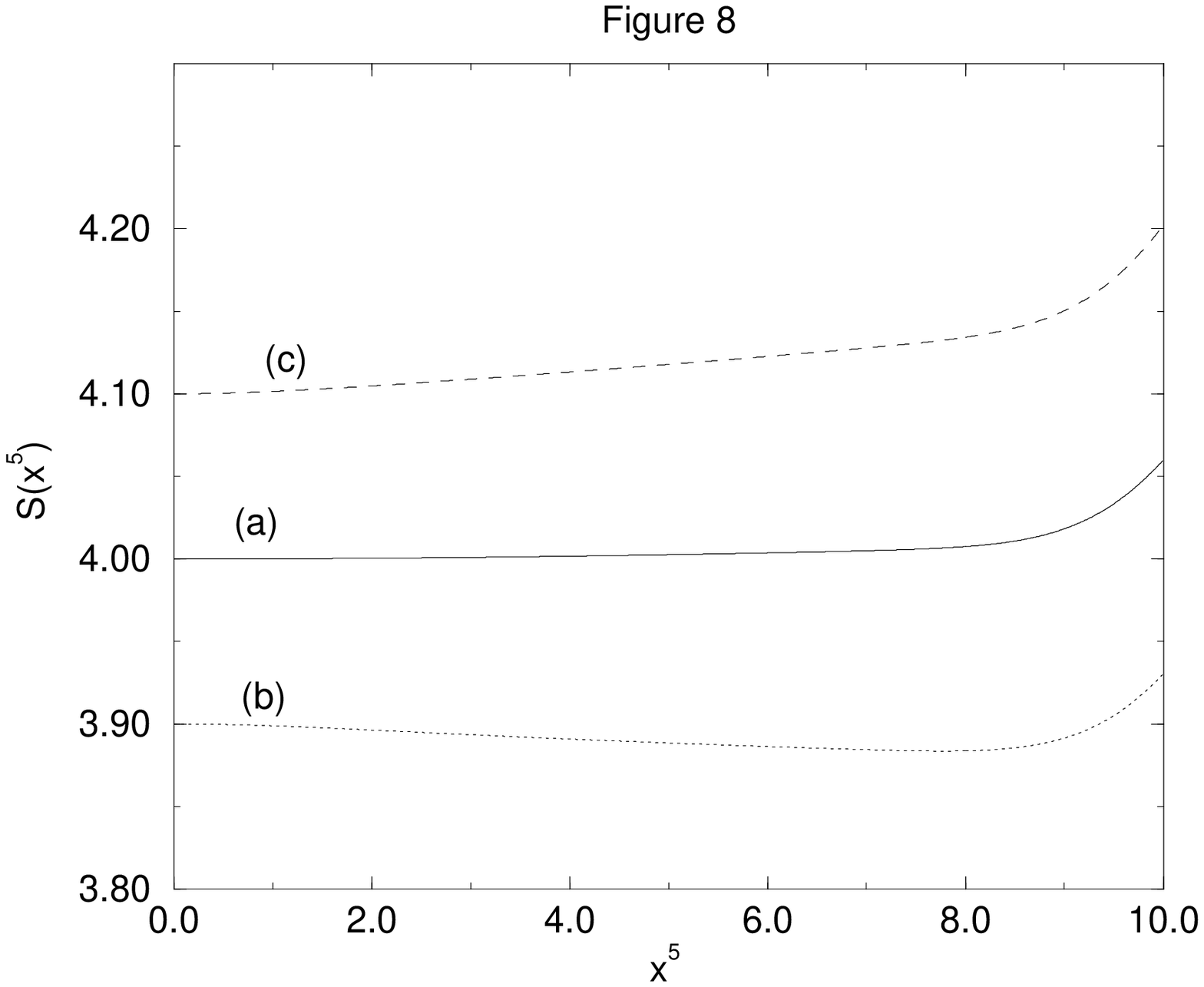,height=3.0in,width=3.0in} 
\psfig{figure=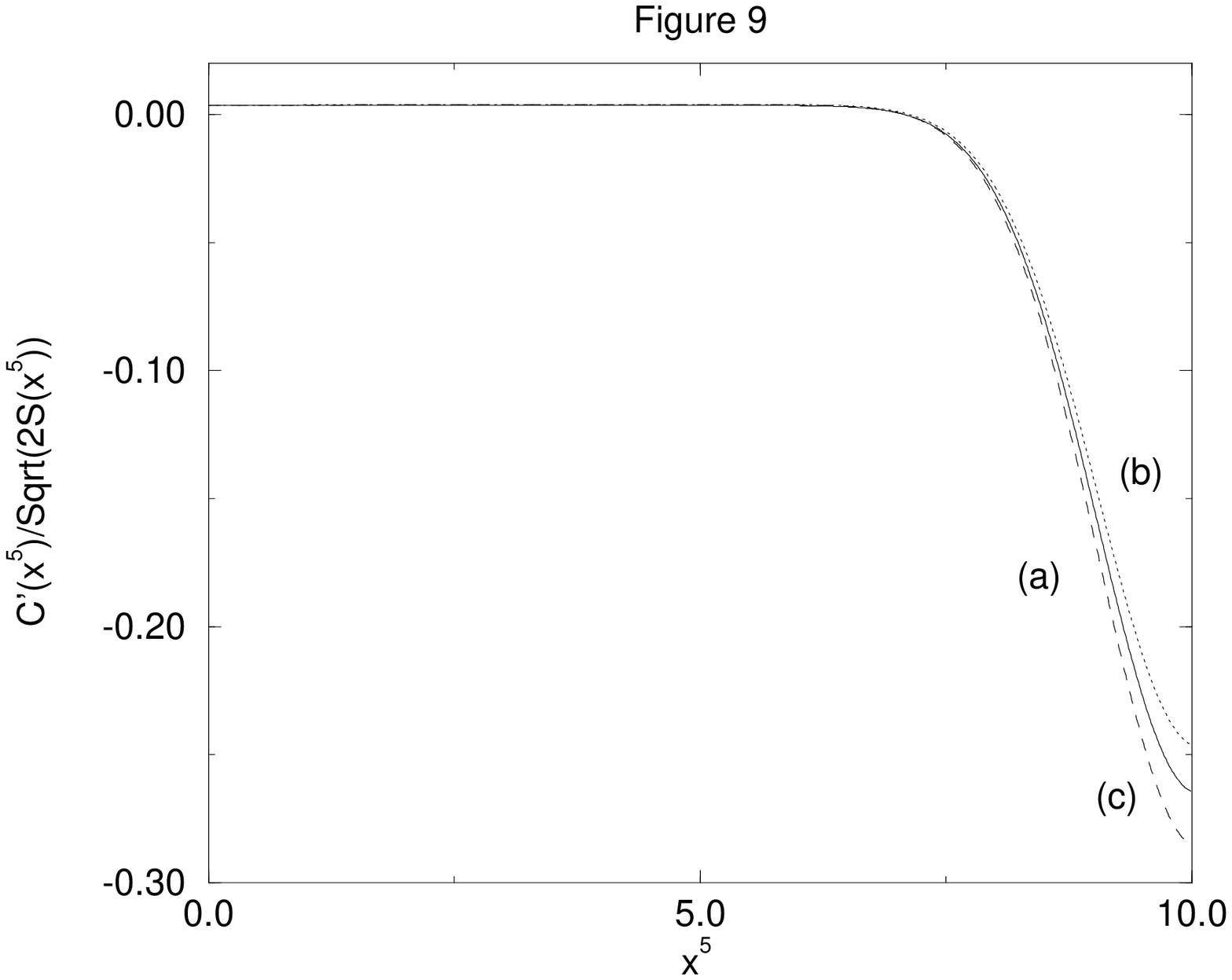,height=3.0in,width=3.0in} 
}} 
\par 
{\small Figure 8: {\em The $x^5$ dependence of the modulus $S(x^5)$ 
in three cases corresponding to (a) - $S(0)=4.0$, (b) - $S(0)=3.9$, 
and (c) - $S(0)=4.1$, cf. Fig. 7.}\\  
 
Figure 9: {\em The $x^5$ dependence of $\frac{C'(x^5=10)}{\sqrt{2 S(x^5=10)}}$ 
for the three cases given in Fig. 8.}} \\ 
\vspace{0.3cm} 
 
Finally, we consider the possibility, very relevant to the possible 
dynamical stabilization of moduli  
in the low-energy effective Lagrangian for $M$ theory, that gauge or other  
non-perturbative dynamics creates a potential for even moduli living on 
the walls, 
and that this potential has two disconnected pieces living on both walls.  
For simplicity, we take the model with just one $Z_2$-even modulus $S$ 
and one $Z_2$-odd field $C$. The modified equations of motion are 
\beqa 
S''(x^{5})&+&\frac{1}{(2 C(x^{5})^2 - S(x^{5}))}  ( -2\, \gamma_{hid}^2 \, 
\delta(x^{5}-\pi \rho)^2 \, 
       ( S_{0,hid} - S(x^{5}) )^3 \, 
       ( 16\, C(x^{5})^4  \nonumber \\ 
&-&      2\, C(x^{5})^2 \, 
          ( S_{0,hid} - 9\,S(x^{5}) )   
    S(x^{5}) \, ( S_{0,hid} + 3 \, S(x^{5}) )  )   \nonumber \\ 
&+& 2 \,( -2\, 
          C(x^{5})^2 + S(x^{5}) ) \, 
       C'(x^{5})^2 - 4\,C(x^{5})\,C'(x^{5})\,S'(x^{5}) + S'(x^{5})^2 \nonumber \\ 
& + &\delta(x^{5}- \pi \rho ) \, 
       ( 8 \, \gamma_{hid} \, 
          ( S_{0,hid} - S(x^{5}) ) \, 
          ( 2\, C(x^{5})^2 + S(x^{5}) ) 
               ^2 \, C'(x^{5})   \nonumber \\ 
&-&  
         8  \, \gamma_{hid} \, C(x^{5}) \, 
          ( S_{0,hid} - S(x^{5}) ) \, 
          ( 2\,C(x^{5})^2 + S(x^{5}) ) 
            \,S'(x^{5}) )) \nonumber \\ 
&+&\frac{1}{(2 C(x^{5})^2 - S(x^{5}))}  ( -2\, \gamma_{vis}^2 \, 
\delta(x^{5})^2 \, 
       ( S_{0.vis} - S(x^{5}) )^3 \, 
       ( 16\, C(x^{5})^4  \nonumber \\ 
&-&      2\, C(x^{5})^2 \, 
          ( S_{0.vis} - 9\,S(x^{5}) )   
    S(x^{5}) \, ( S_{0.vis} + 3 \, S(x^{5}) )  )   \nonumber \\ 
& + &\delta(x^{5}) \, 
       ( 8 \, \gamma_{vis} \, 
          ( S_{0.vis} - S(x^{5}) ) \, 
          ( 2\, C(x^{5})^2 + S(x^{5}) ) 
               ^2 \, C'(x^{5})   \nonumber \\ 
&-&  
         8  \, \gamma_{vis} \, C(x^{5}) \, 
          ( S_{0.vis} - S(x^{5}) ) \, 
          ( 2\,C(x^{5})^2 + S(x^{5}) ) 
            \,S'(x^{5}) )) = 0 
\eeqa 
\beqa 
C''(x^{5})&-&\gamma_{hid}\, (S_{0,hid} -S(x^{5}))^2 \delta'(x^{5}-\pi \rho) 
+ \frac{1}{(2 C(x^{5})^2 - S(x^{5}))}( 4\,\gamma_{hid}^2\,C(x^{5})\, \nonumber \\ 
& &\delta(x^{5}-\pi \rho)^2\,( S_{0,hid} - 4\,C(x^{5})^2 - 3\,S(x^{5}) ) \, \nonumber \\ 
& & ( S_{0,hid} - S(x^{5}) ) ^3   
- 4\,C(x^{5})\,C'(x^{5})^2 +  2\, \gamma_{hid}\,\delta(x^{5}-  
\pi \rho)\, \nonumber \\ 
& & ( S_{0,hid} - S(x^{5}) ) \, 
   ( 2\,C(x^{5})^2 + S(x^{5}) ) \, 
   ( 4\,C(x^{5})\,C'(x^{5}) -  
     S'(x^{5}) ) \nonumber \\ 
&+& C'(x^{5})\,S'(x^{5}))  
- \gamma_{vis} (S_{0,vis} -S(x^{5}))^2 \delta'(x^{5}-\pi \rho) \nonumber  \\ 
&+& \frac{1}{(2 C(x^{5})^2 - S(x^{5}))}( 4\,\gamma_{vis}^2\,C(x^{5})\, 
   \delta(x^{5}-\pi \rho)^2\, \nonumber \\ 
& &( S_{0,vis} -  
     4\,C(x^{5})^2 - 3\,S(x^{5}) ) \, 
   ( S_{0,vis} - S(x^{5}) ) ^3 \nonumber  \\ 
&+& 2\, \gamma_{vis}\,\delta(x^{5}- \pi \rho)\,( S_{0,vis} - S(x^{5}) ) \, \nonumber \\ 
& &   ( 2\,C(x^{5})^2 + S(x^{5}) ) \, 
   ( 4\,C(x^{5})\,C'(x^{5}) -  
     S'(x^{5}) ) ) = 0 
\eeqa 
The two-piece potential for $S$ that we assume, expanding 
each part of the effective potential to quadratic order around its respective   
`naive' minimum, is  
\beq 
V_{eff}(S)= \gamma_{vis} (S-S_{0,vis})^2 \delta (x^5) +  
\gamma_{hid} (S-S_{0,hid})^2 \delta (x^5-\pi \rho) 
\eeq 
We assume that each part of the potential has a local minimum,  
taken to be $S_{0,vis}=4.0$ on the visible wall and $S_{0,hid}=1.0$ on the  
hidden wall. The coefficients $\gamma_{vis,hid}$ are taken to be equal to  
$-0.2$ on both walls.  
In such a situation, when the coefficients of  
$\delta$-like 
sources depend on boundary configurations of fields which are also subject  
to bulk dynamics, one expects that both boundary values and bulk  
configuration are to be chosen in a self-consistent way.  
 
To understand the situation  
better, we have computed numerically the vacuum expectation value of the 
total Hamiltonian 
of the system (excluding gravity) as a function of boundary conditions on the  
visible wall. We see in Fig.~7 that the local  
minimum of the Hamiltonian corresponds to the local minimum of the part of the  
potential living on the visible wall. However, this minimum value is 
non-zero,  
signalling an inhomogeneous configuration in the bulk, and, as seen  
from Fig.~8, it gives boundary values on the second wall that are  
far away from the local minimum of the second part of the potential for $S$, 
which is naively $S_{0,hid}=1.0$. Of course, the reasoning can be reversed 
with  
respect to the walls, and there exists another local minimum of the 
Hamiltonian, 
where the boundary value of $S$ on the hidden wall is close to $S_{0,hid}=1.0$ 
but differs from $S_{0,vis}=4.0$ on the observable wall\footnote{The local  
minimum  
corresponding to the configuration where $S_{0,vis}=4.0$ gets cancelled  
is the deeper of the two.}. 
Hence, the existence of the two disconnected pieces of the moduli potential  
(which may or may not be related to gaugino condensation on the walls) 
plus the bulk dynamics lead to  
the interesting phenomenon of an apparently `displaced' moduli vacuum 
on one of the walls, which is a sign 
of the existence of the second wall.  
 
We have discussed in this Section  
supersymmetry breaking at the level  
of the five-dimensional supergravity. The expectation values of  
variations of even components of bulk  
fermions which do not depend on fermionic fields and do not contain 
four-dimensional space  
derivatives are identified as possible signatures and measures of 
supersymmetry  
breaking. These are bulk variations, but they are continuous across the walls, 
as are the even fermionic fields, so they constitute legal 
supermultiplets  
living on the four-dimensional walls. Similarly to what has been observed 
in the example of a globally supersymmetric toy model with 
vector  
multiplets~\cite{peskin}, the terms we have identified in the  
supersymmetric variations of fermions are to be interpreted from the point 
of view of  
four-dimensional boundary chiral multiplets as gravitational contributions  
to the $F$ terms of chiral superfields. These contributions are 
gravitational in the sense that they are 
suppressed by additional powers of the $m_{11}$ scale, namely 
$\kappa^{2/3}$, 
as these terms contain in our example 
a factor of $\kappa^{2/3}$ due to the gauge coupling to the source 
on the hidden wall (it is implicit in $\cs$). In the five-dimensional canonical  
frame  
they are also inversely proportional  
to the distance $\pi \rho$ 
between the walls, as in (\ref{oef}), (\ref{der66}) and the discussion  
in Section 3: 
\beq 
\delta F_{S,T} = \alpha_{S,T} \frac{\cs}{\pi \rho \sqrt{V_{CY}}}    
\eeq 
where $\alpha_{S,T}$ is a coefficient of order one depending on whether  
we look at supersymmetry variations of hyperinos ($S$) or  
$\psi_5$ ($T$), as seen in (\ref{psif}), (\ref{beef}). 
 
Before attempting to identify the soft terms breaking  
global supersymmetry on the walls, one  
should take care of some subtleties. First, although the superfields we 
discuss 
here look like true chiral multiplets on the wall, one has to remember that  
they come from the gravitational bulk sector.  
The basis we work with in the bulk gives the canonical Einstein-Hilbert  
and gravitino action in five dimensions, {\em not} in four. When going 
over to four dimensions, one has to perform field  
and metric redefinitions on the walls, to obtain the field frame  
which is canonical in four dimensions. The necessary ingredient is then the  
specific definition of the effective four-dimensional gravitational 
sector.  
As we have learned from the toy models discussed in Section 3,  
the proper way to 
define effective moduli-charged matter couplings is to compute them on the  
wall as limiting values of solutions to the equations of  
motion along the fifth dimension. In principle, gravitational degrees of 
freedom  
are no exception, and the same procedure as the one applied in Section 3  
to a general $Z_2$-even field should be carried out.  
Hence,  
one can carry out the  
programme of Weyl rescaling the metric, 
separating out the component $\psi^{+}_{5}$ and redefining the 
four-dimensional gravitino 
in order to produce a canonical kinetic term for $\psi^{+}_{5}$, and 
correcting  
suitably the supersymmetry variations to account for  
redefinitions~\cite{bergshoff}. The details of this construction of the  
four-dimensional effective action demand 
a separate discussion in themselves, and we do not attempt  
puruse them in  
the present work. In the present paper we have worked 
in the canonical five-dimensional basis, and in this basis the goldstino on the  
wall is a mixture of the $Z_2$-even fermion from the $T$-plet and 
hyperinos from the $S$ multiplet and from non-universal hypermultiplets.

Finally, we comment further on some of the questions 
discussed earlier, in particular on 
the transmission of supersymmetry breaking between the walls,  
in the light of the corrections to the effective action that arise in 
the gauged five-dimensional supergravity construction of~\cite{low5}. 
We recall briefly that the gauging arises because the vacuum 
solution has non-vanishing 
components of the antisymmetric tensor field and its strength, 
linear in $x^{11}$ to lowest non-trivial  
order in $\kappa^{2/3}$. Hence, in the construction of the  
effective five-dimensional theory, one expands the Lagrangian around 
this non-trivial eleven-dimensional  
background, and treats five-dimensional zero modes as fluctuations in 
that non-vanishing  
background~\cite{low5}. Upon substituting such an expansion into the 
topological $C \wedge G \wedge G $ term in the  
eleven-dimensional supergravity Lagrangian,  
one finds, among other terms, a new coupling between zero modes  
of the form $\partial_{\mu} D \; A^{\mu}$, where $D$ is in our language 
the  
imaginary part of the complex even scalar $S$ from the universal 
hypermultiplet, and in the language of the effective 4d theory on a wall 
is simply the universal axion. The vector boson $A$ is a  
particular combination of the $h_{1,1}$ Abelian vector bosons existing  
in the five-dimensional bulk, whose composition depends on the 
orientation of  
the gauge and gravitational instantons with respect to the cohomology basis  
used to define the zero modes. Note that this term 
is of order ${\cal O}(\kappa^{2/3})$, which is of higher order 
than the kinetic couplings in the bulk which we have considered  
up till now in  
the present paper. This new term breaks the zeroth-order supersymmetry in 
the bulk, and should be supersymmetrized 
when one works consistently with the higher-order bulk couplings. 
The unique way to supersymmetrize such a  
derivative scalar-vector coupling is to regard it as a part of a  
covariant derivative in a gauged supergravity model~\cite{low5}. The 
particular form of that coupling enables one to identify the Killing 
vector of the quaternionic manifold isometry which is gauged, and  
formulae given in~\cite{andr} can be used to read off the remaining terms 
in the Lagrangian that are needed to maintain supersymmetry, and the 
corresponding modifications to the supersymmetry  
transformation laws. The required terms  
in the Lagrangian were obtained in~\cite{low5} from the   
reduction of the eleven-dimensional Lagrangian with a non-vanishing 
background. One particularly interesting new term related by 
supersymmetry to the 
${\cal O}(\kappa^{2/3})$ term is a potential term, analogous  
to $D$ terms in four-dimensional supersymmetry, which is of 
order ${\cal O}(\kappa^{4/3})$. Thus, the gauged theory contains new 
couplings  
in the bulk, between fields belonging to different multiplets,  
but they are of higher order in an expansion in powers of $\kappa^{2/3}$ 
than the couplings we have considered up till now.  
 
We note that the coupling to bulk fields of a source on the wall, 
such as a gaugino condensate, is already 
suppressed by a power of $\kappa^{2/3}$. The effects on the 
transmission of supersymmetry breaking of new terms in the bulk  
are of higher order, and hence 
unlikely to change qualitatively the conclusions we have reached 
working  
with the leading-order Lagrangian: they would simply contribute  
additional mixing of the scalars and vectors living in the bulk. 
One must further check the 
supersymmetry transformations laws, which are also modified. However, as 
as has been already noticed in~\cite{low5}, the corrections to the 
transformations, which are linear in the non-trivial background, are 
not only of higher order, as is that background, but also discontinous 
across the walls, 
since the background to which they are proportional is  
itself discontinuous.  
This means that these corrections do not appear on the walls, 
and so do not open up any new 
channels of communication of supersymmetry breaking  
to the fields living on the visible wall, beyond  
those already identified in the present paper. 
 
Finally, we observe that the origin of non-trivial 
backgrounds of certain five-dimensional zero modes, such as the real part 
of $S$ which represents the Calabi-Yau volume, is traceable 
to non-trivial sources living  
on the walls. Both in the case discussed here 
and in the gauged supergravity model, these are 
coupled to zero modes that change  
quasi-linearly across the bulk. The role of such sources, 
which we have studied  
in this paper in the leading-order Lagrangian,  
continues to hold to leading order also in the presence of the terms 
associated with the gauging, as do our conclusions.  
 
To summarize this discussion, the gauging of the five-dimensional  
supergravity induced by a non-trivial background  
for the antisymmetric tensor field in ${\cal O}(\kappa^{2/3})$ 
induces higher-order corrections  
to our results, due to additional higher-order  mixing between 
bosonic fields in the bulk. However, our conclusions on the  
possible patterns of transmission of supersymmetry breaking remain 
unaffected by the gauging.

\section{Conclusions} 
 
We have argued in this paper that a systematic reduction of the 
eleven-dimensional  
$M$-theory effective Lagrangian should always proceed in two steps. 
First, the reduction from eleven to five dimensions should be performed.  
This yields a five-dimensional Lagrangian that 
is richer in parameters than in eleven dimensions, due to effects related 
to 
the geometry of the compact Calabi-Yau manifold. In 
five dimensions, one 
always  
obtains a non-linear $\sigma$ model for bulk moduli fields, which implies  
non-trivial dynamics across the $Z_2$ orbifold. We have presented  
simple examples of such models, and have analyzed with their help the  
evolution of the moduli between the walls. In the examples discussed, which  
belong to the class of ungauged five-dimensional supergravities, the 
non-trivial  
configurations in the bulk have to be excited by sources living on the  
four-dimensional walls.  
 
There exist  
natural sources which may come from the gauge models living on the 
boundaries. A favoured 
example is a condensate of gauge fermions in a strongly-coupled  
gauge group living on the opposite end of the fifth dimension 
from the visible four-dimensional sector. The 
configurations induced by a non-vanishing  
condensate lead, upon solving the equations of motion in the fifth 
dimension,  
to non-vanishing supersymmetry variations of modulini at 
the second, visible wall. The relevant variations are computed locally  
on the visible wall, and are inversely proportional to the square root of 
the  
Calabi-Yau volume, and directly proportional to vacuum expectation values 
of derivatives  
with respect to the fifth coordinate of the $Z_2$-odd moduli living in the  
bulk. We have also given a general argument that the 
effective strength of  
the supersymmetry breaking operators induced this way on the visible wall  
decreases as the inverse of the distance between the walls. The correlations  
between the magnitude of the visible supersymmetry breaking and the  
scale of the hidden condensate, and between some other characteristics  
of the dynamics across the fifth dimension, have been presented in a model  
containing just a single universal supermultiplet, and in the second 
case, one which also has one non-universal supermultiplet. 
One interesting observation concerns the case where a stiff condensate 
is replaced by effective boundary superpotentials. Then the local vacuum  
on the observable wall can be visibly different from the naive one 
computed from the effective moduli potential born directly on the visible 
wall.  
The shift of the minima for the moduli is due to the component of 
the potential on the second, hidden, wall, transmitted by the 
equations of motion through the bulk.  
 
Our findings are not in disagreement with  
the eleven-dimensional discussion of~\cite{horava}. We 
find that when the condensate is switched on, it induces 
supersymmetry breaking locally, which cannot be removed by a legal 
redefinition of the 
parameter of the supersymmetry transformation. This acts as a source 
for the $\sigma$ model in the bulk, and the solution of the bulk equations 
of motion in the presence of 
this source is such that the resulting boundary  
values of the moduli and their derivatives on the observable wall also  
correspond to a breaking of global supersymmetry restricted to this wall.  
 
In this paper, we have barely scratched the surface of the rich physics 
opened up by the five-dimensional framework~\cite{ccf,aft,low5} explored 
here. There are 
many fascinating open issues such as the exact origin of the 
supersymmetry-breaking sources on the hidden wall, details of the  
coupling of the walls to the bulk theory, the  
effects  of the deformation of the 
Calabi-Yau manifold in the model discussed here, the appearance of 
non-trivial dynamics for the complex structure moduli, the dynamical 
choice of the vacuum and the magnitude of the fifth dimension,  
applications to specific `realistic' Calabi-Yau compactifications, the 
possible extension to non-Calabi-Yau reductions from eleven to five 
dimensions, and many more. We can hope that some of the unresolved 
issues plaguing weak-coupling string models may by illuminated by 
five-dimensional light.\\ 
 
\noindent{\bf Acknowledgments}: 
 
Z.L. and S.P. are supported in part 
by the Polish Commitee for Scientific Research grant 2 P03B 040 12, and  
by the M. Curie-Sklodowska Foundation. W.P. would like to thank Professor 
Graham G. Ross for stimulating discussions.\\

\end{document}